\documentclass[runningheads]{llncs}
\usepackage[T1]{fontenc}
\usepackage{graphicx}
\usepackage[nolist,nohyperlinks]{acronym} 
\acrodef{Acc}{Accurate model}
\acrodef{AI}{Artificial Intelligence}
\acrodef{AUC}{Area Under the receiver operating characteristic Curve}
\acrodef{CBM}{Concept Bottleneck Model}
\acrodef{CExp}{Concept Explanation}
\acrodef{CExp+Int}{Concept Explanation with Interventions}
\acrodef{CExp+Int+SMap}{Concept Explanation with Interventions and Saliency maps}
\acrodef{CM}{Concept Model}
\acrodef{CUB}{Caltech-UCSD Birds-200-2011}
\acrodef{DDI}{Diverse Dermatology Images}
\acrodef{DNN}{Deep Neural Network}
\acrodef{Inacc}{Inaccurate model}
\acrodef{NoExp}{No Explanations}
\acrodef{NoInt}{No Interventions}
\acrodef{LR}{Learning Rate}
\acrodef{ML}{Machine Learning}
\acrodef{SCS}{System Causability Scale}
\acrodef{SGD}{Standard Gradient Descent}
\acrodef{SUS}{System Usability Scale}
\acrodef{SGD}{Stochastic Gradient Descent}
\acrodef{WithInt}{With Interventions}
\acrodef{XAI}{eXplainable Artificial Intelligence}

\usepackage{siunitx}
\usepackage{makecell}
\usepackage[caption=false]{subfig}
\usepackage{tcolorbox}

\begin{document}
\title{The Impact of Concept Explanations and Interventions on Human-Machine Collaboration}
\titlerunning{The Impact of Concept Explanations and Interventions}
%
\author{Jack Furby\inst{1}\orcidID{0000-0002-2348-8091} \and
Dan Cunnington\inst{2} \orcidID{0000-0003-0715-964X}\and
Dave Braines\inst{3}\orcidID{0000-0003-3296-0842} \and
Alun Preece\inst{1}\orcidID{0000-0003-0349-9057}}
\authorrunning{J. Furby et al.}
%
\institute{Cardiff University, UK \and
Imperial College London, UK \and
IBM Research Europe}
\maketitle              
\begin{abstract}
Deep Neural Networks (DNNs) are often considered black boxes due to their opaque decision-making processes. To reduce their opacity Concept Models (CMs), such as Concept Bottleneck Models (CBMs), were introduced to predict human-defined concepts as an intermediate step before predicting task labels. This enhances the interpretability of DNNs. In a human-machine setting greater interpretability enables humans to improve their understanding and build trust in a DNN. In the introduction of CBMs, the models demonstrated increased task accuracy as incorrect concept predictions were replaced with their ground truth values, known as intervening on the concept predictions. In a collaborative setting, if the model task accuracy improves from interventions, trust in a model and the human-machine task accuracy may increase. However, the result showing an increase in model task accuracy was produced without human evaluation and thus it remains unknown if the findings can be applied in a collaborative setting. In this paper, we ran the first human studies using CBMs to evaluate their human interaction in collaborative task settings. Our findings show that CBMs improve interpretability compared to standard DNNs, leading to increased human-machine alignment. However, this increased alignment did not translate to a significant increase in task accuracy. Understanding the model’s decision-making process required multiple interactions, and misalignment between the model’s and human decision-making processes could undermine interpretability and model effectiveness.

\end{abstract}

\section{Introduction} \label{intro}

\acfp{CM}, such as \acp{CBM} \cite{concept_bottleneck_models}, is a class of \ac{DNN} which aims to improve the interpretability of model predictions by structuring predictions around human-understandable components, called concepts. These concepts often correspond to intermediate attributes of tasks, effectively ``splitting'' the prediction process into sub-tasks. For instance, a \ac{CM} that predicts types of birds, might predict concepts for the wing colour and beak shape.

After a \ac{CM} makes a prediction a human collaborating with the model will be able to inspect concept predictions, known as \textit{concept explanations}, to help understand the model's decision-making process. In domains such as healthcare this may be used to answer why a downstream task was predicted. With some \ac{CM} architectures, such as \acp{CBM}, the concept explanations also introduce the capability for a human to intervene in the concept predictions. As \acp{CBM} predict concepts in the range of 0 to 1 (not present to present) with a 0.5 threshold, concept outputs can be replaced with new values within this range. Interventions may be made to correct mistakes the model made when predicting concepts, or otherwise query the model to see what task prediction would be made with a different set of concept predictions. In these scenarios the model's concept vector provides counterfactual explanations \cite{concept_bottleneck_models}. In a collaborative setting, we may consider interventions as a way to increase trust in a model as interventions may reveal the sensitivity a model has to concept values, and thus any bias the model has to concepts.

The authors of \acp{CBM} \cite{concept_bottleneck_models} presented results using automated metrics demonstrating improved model accuracy as incorrect concept predictions were intervened and replace with their ground truth values.  In addition, model predictions have been shown as a preferred method to identify bias \cite{10.5555/3495724.3495784}. As the intervention metric was automated it remains unknown if the findings can be applied in a collaborative setting.

The authors of \acp{CBM} also made claims of improved human-machine collaboration \cite{concept_bottleneck_models}, but human studies to show this are limited and instead compare \acp{CBM} to other model architectures \cite{x-char,dubey2022scalable}, or complete tasks such as selecting the concepts participants believe the model detected \cite{Learning_Bottleneck_Concepts}. Only a few studies analyse the class of \acp{CM} with collaborative tasks \cite{Editable_User_Profiles,humans_guide_machines_xai}.

This paper presents two human studies where we analysed \acp{CBM} in a collaborative setting. We answer the research question: \textit{Do Concept Models improve task accuracy and model interpretability in a human-machine setting?} We have broken this question down into the following sub-questions:

\begin{enumerate}
    \item Do test-time interventions improve human-machine task and concept accuracy?
    \item Do interventions increase the interpretability of \acp{CBM}?
    \item Are \acp{CBM} trusted?
\end{enumerate}

The main contributions of this paper are as follows:

\begin{itemize}
    \item We perform the first human studies using \acp{CBM} in a joint human-machine task setting which analyses the interaction between humans and the \ac{CBM}. We find interventions often increased trust in a model but this trust was sometimes misplaced. In addition, the \ac{CBM} decision-making process is not aligned with that of the humans.
    \item We show the initial promise of interpretability from high-level concepts is upheld. However, understanding the model's decision-making process requires participants to actively interact with the model. Additionally, providing concept predictions without the capability to intervene has a similar effect on task accuracy.
\end{itemize}

Although we used \acp{CBM}, our findings also apply to other \acp{CM} with intervention capabilities and that predict task labels in the same feed-forward fashion from input to concepts, to task label. Namely these are Concept Embedding Models \cite{cem}, Sidecar \acp{CBM} \cite{sidecar_cbm}, and hybrid \acp{CBM} \cite{Mahinpei2021PromisesAP}.

\section{Related Work}

Several studies have analysed \acp{CM} with human participants. These can be placed into several categories; human concept preference \cite{Barker2023SelectiveCM,Ramaswamy_2023_CVPR}, concept explanations \cite{x-char,jeyakumar2022automatic,sixt2022do,dubey2022scalable}, human-in-the-loop \cite{Editable_User_Profiles,humans_guide_machines_xai} and bias discovery \cite{yuksekgonul2022posthoc,posthoc_cbm_repeat}. 

In studies on concept preference, one study found that concepts humans identified in samples varied widely and performed worse when used by a \ac{CM} on downstream tasks compared to those identified by the model \cite{Barker2023SelectiveCM}. Separately, Participants have also been found to prefer to see 32 or fewer concepts \cite{Ramaswamy_2023_CVPR} instead of all concepts a model uses for task predictions, if greater than 32. This is consistent with balancing a models completeness \cite{6645235} to keep participants engaged \cite{10.1145/2678025.2701399}.

Next, studies exploring concept explanations preference have demonstrated a mixed response where concept explanations are favoured in some studies \cite{jeyakumar2022automatic}, while not in others \cite{sixt2022do,dubey2022scalable,x-char}. Out of these studies \cite{sixt2022do,dubey2022scalable} evaluated explanation types for bias discovery and model task prediction. As concept explanations underperformed in this area it raises the question of whether \ac{CM}'s concept explanations improve the interpretability of models such that they can aid in human-machine collaboration.

Some studies have investigated human-machine collaboration with \acp{CM}. A \ac{CBM} inspired recommender system was introduced that combined user-provided and automatically generated concepts to suggest relevant text documents \cite{Editable_User_Profiles}. Participants could intervene in the concepts by editing a concept value, leading to improvements of 20–47\% in recommendation accuracy compared to initial concept values. A separate study asked participants to label images \cite{humans_guide_machines_xai}. Participants either had fixed model predictions to help them or could select parts of the input image for the model to focus on, finding little difference in performance (73.57\% vs. 72.68\% respectively). Additionally, participants often agreed with the model’s predictions regardless of whether the model was correct or incorrect. Both of these studies look at humans updating a model's prediction, similar to interventions with \acp{CBM}. As the recommender system \cite{Editable_User_Profiles} is explicitly based on \acp{CBM}, their findings suggest similarities could be observed in an image modality.

For bias discovery, \cite{yuksekgonul2022posthoc} (and repeated in the study \cite{posthoc_cbm_repeat}) used \ac{CBM}-like architectures to study human-guided pruning on a model where input samples had shifted (e.g. the correlation of concepts co-occurring is changed after training). Participants selected concepts to prune based on input samples and model predictions, outperforming random pruning and only slightly less effective than fine-tuning or greedy performance. This demonstrates that the concept explanations are effective at aiding a human-in-the-loop understanding of bias in a model.

From these studies, we have identified no papers which look at \acp{CBM} or similar model architectures that evaluate the capabilities of \acp{CBM} in real-world tasks. Most importantly it has not been shown if human performed interventions improve joint task performance, and whether these models are more interpretable than standard \acp{DNN}.

\section{Methods}

We ran two human studies: (1) An expert study where participants had extensive knowledge about the task domain (skin disease diagnosis) where the model acted as a second opinion. (2) A lay-person study with a general task (Playing games of Blackjack), involving participants with experience levels ranging from novices to skilled individuals, but none being professionals. The model also acted as a second opinion, but could also serve as a guide for participants with less experience. Both studies received favourable ethical opinion from the School of Computer Science and Informatics at Cardiff University.

By running two studies we were able to compare results in similar, but distinct settings where participants can interact with an \ac{AI} agent to assist them. Following the taxonomy by \cite{rigorous_xai}, our lay-person study is human-grounded as we do not use expert participants and use a simulated task, while the expert study is application-grounded as we use both expert participants and a real-world task.

\begin{table}[t]
    \centering
    \caption{Participants in the expert study were split into two groups, both with access to the same model. Participants in the lay-person study were split into eight groups where the model used and explanations provided were varied.}
    \begin{tabular}{l|ll}
    \hline
    Expert study                         & \multicolumn{2}{l}{Lay-person study}                                                             \\ \hline
    Participant groups                   & \multicolumn{2}{l}{Participant groups}                                                           \\ \hline
                                         & \multicolumn{1}{l|}{Accurate model}                       & Inaccurate model                     \\ \cline{2-3} 
    \acs{CExp+Int}      & \multicolumn{1}{l|}{\acs{NoExp}}         & \acs{NoExp}         \\
    \acs{CExp+Int+SMap} & \multicolumn{1}{l|}{\acs{CExp}}          & \acs{CExp}          \\
                                         & \multicolumn{1}{l|}{\acs{CExp+Int}}      & \acs{CExp+Int}      \\
                                         & \multicolumn{1}{l|}{\acs{CExp+Int+SMap}} & \acs{CExp+Int+SMap} \\ \hline
    \end{tabular}
    \label{tab:group_breakdown}
\end{table}

\begin{figure}[t]
    \centering
    \subfloat[Only task predictions (\acs{NoExp})\newline\newline]{
        \includegraphics[width=0.22\textwidth]{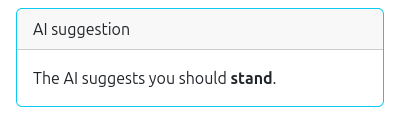}
        \label{fig:blackjack_no_explanation_version_concepts}
    }
    \hfill
    \subfloat[Task predictions and top three concept predictions (\acs{CExp})\newline]{
        \includegraphics[width=0.22\textwidth]{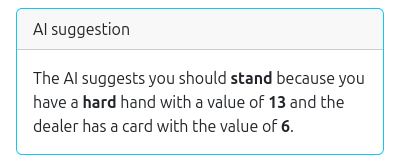}
        \label{fig:blackjack_explanation_version_just_concepts}
    }
    \hfill
    \subfloat[Task predictions, concept predictions and interventions (\acs{CExp+Int})\newline]{
        \includegraphics[width=0.22\textwidth]{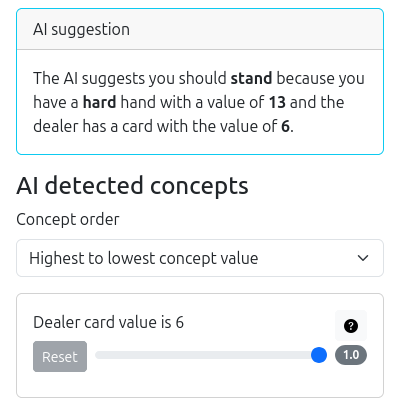}
        \label{fig:blackjack_explanation_version_concepts}
    }
    \hfill
    \subfloat[Task predictions, concept predictions, interventions and saliency maps (\acs{CExp+Int+SMap})]{
        \includegraphics[width=0.22\textwidth]{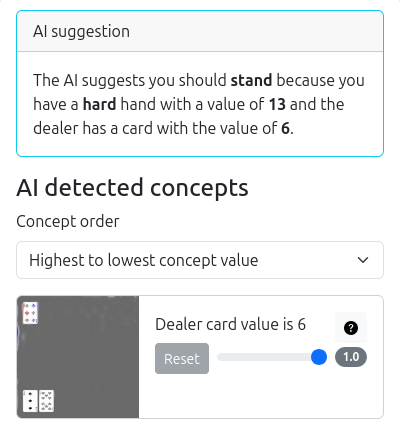}
        \label{fig:blackjack_explanation_version_full}
    }
    \caption{Model output variations.}
    \label{fig:explanation_versions}
\end{figure}

We split participants into two groups in the expert study, and eight groups in the lay-person study, detailed in Table~\ref{tab:group_breakdown}, and example explanations are shown in Figure~\ref{fig:explanation_versions}. As our sub-questions required us to analyse the use of interventions, interpretability, and trust of \acp{CBM}, both groups for the expert study included participant access to interventions. As we did not have the same limitation in the lay-person study we also included groups that had access to a different model and included groups with no model explanations and just concept explanations. 12 participants took part in the expert study who were either doctors, consultants or trainees with expertise in dermatology. 104 participants took part in the lay-person study where most were either university staff or students.

We use the following acronyms to separate each participant group:

\begin{description}
    \item[\acs{Acc}] Accurate model (lay-person study only).
    \item[\acs{Inacc}] Inaccurate model (lay-person study only).
    \item[\acs{NoExp}]  No explanations (lay-person study only).
    \item[\acs{CExp}] Predicted task label and concept explanations (lay-person study only).
    \item[\acs{CExp+Int}] \acs{CExp} plus Intervention capability.
    \item[\acs{CExp+Int+SMap}] \acs{CExp+Int} plus Saliency maps.
    \item[\acs{WithInt}] Participants who performed interventions or samples where interventions were performed.
    \item[\acs{NoInt}] Participants who did not perform interventions or samples where no interventions were performed.
\end{description}

\acs{Acc} and \acs{Inacc} are placed before the model output and feature capabilities. \acs{WithInt} and \acs{NoInt} are placed after model output and feature capabilities. For example, participants using the accurate Blackjack model with concept explanations and interventions, and who performed interventions would be referred to as \acs{Acc}-\acs{CExp+Int}-\acs{WithInt}.

\subsection{Human Study Design}

Expert study participants were asked to diagnose skin conditions in 10 images from the Skincon dataset \cite{skincon}. We excluded images that were out of focus and limited images to those with the label ``malignant melanoma'' and ``seborrhoeic keratosis'' as a dermatologist would typically look to diagnose a patient. The images were shown in a random order.

\begin{figure}[ht!]
    \centering
    \subfloat[Expert study]{
        \includegraphics[width=0.48\textwidth]{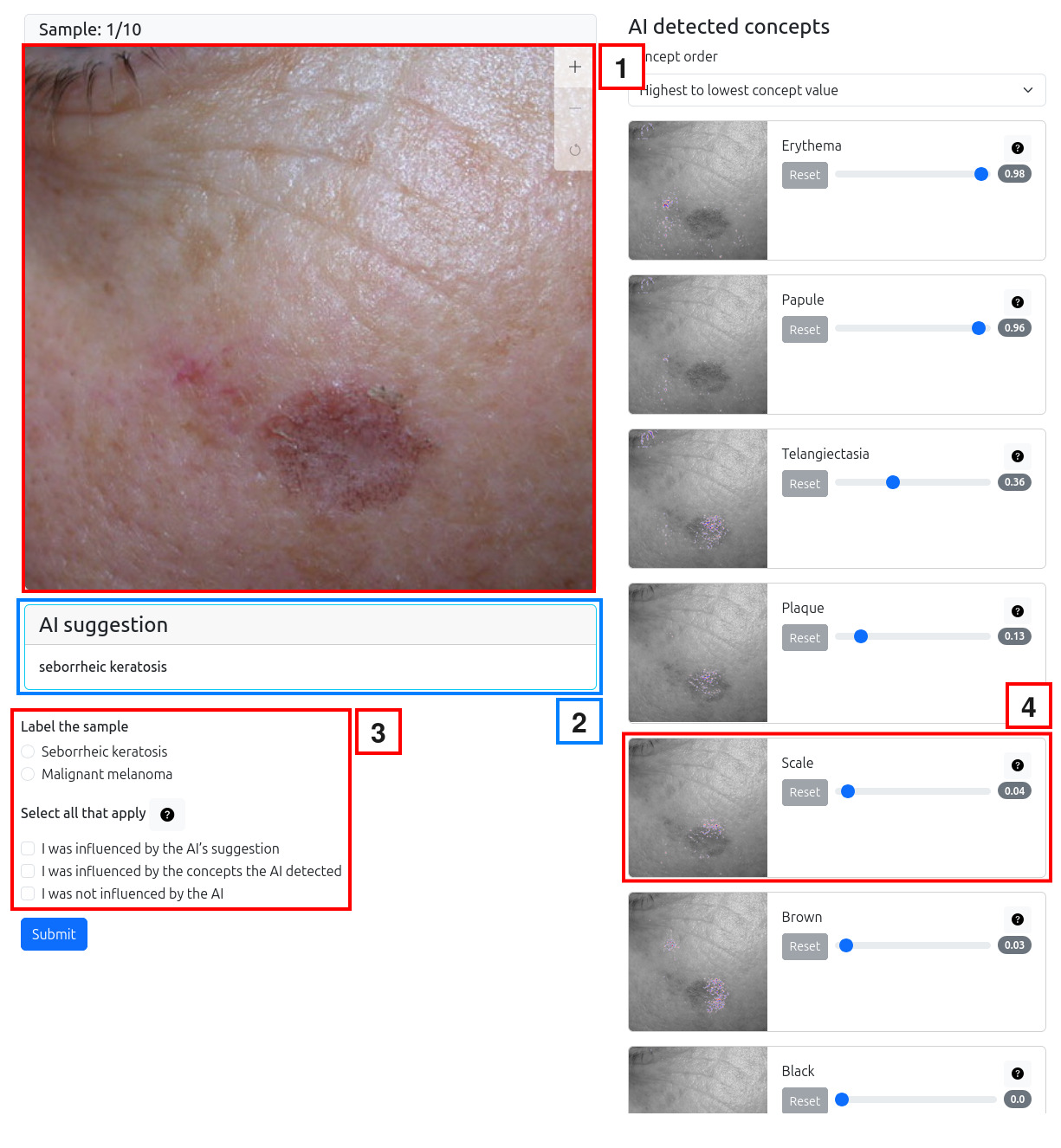}
        \label{fig:skincon_interface}
    }
    \hfill
    \subfloat[Lay-person study]{
        \includegraphics[width=0.48\textwidth]{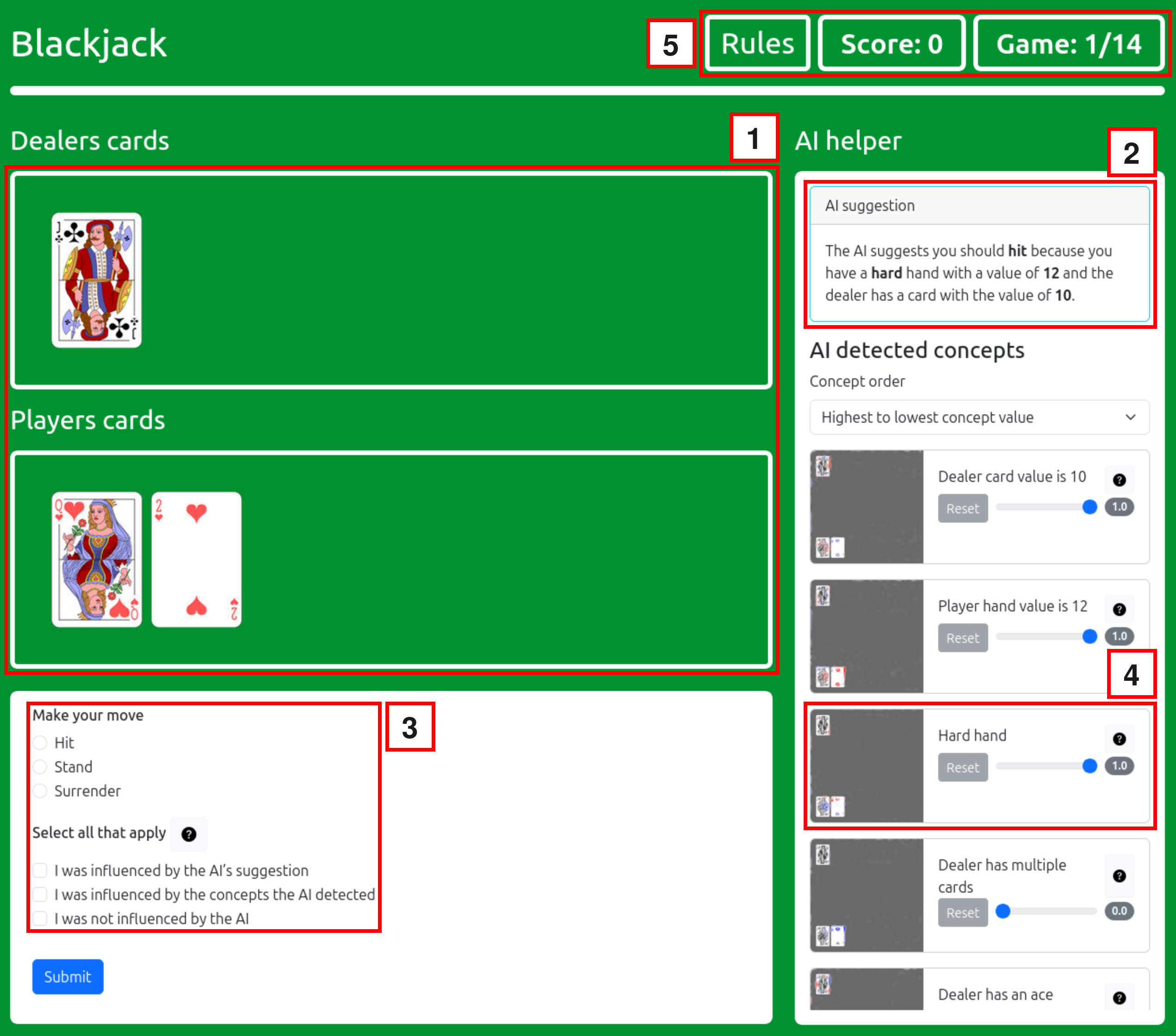}
        \label{fig:blackjack_interfaces}
    }
    \hfill

    \begin{minipage}{\textwidth}
        \begin{tcolorbox}[boxrule=1pt, colframe=black, colback=white]
            \begin{itemize}
                \item[1:] Sample image
                \item[2:] \ac{AI} agent output class and concept labels
                \item[3:] Label selection and \ac{AI} use options for the participant to select
                \item[4:] Concept outputs, salience map, and intervention slider
                \item[5:] Game rules, participant score, and game counter (lay-person study)
            \end{itemize}
        \end{tcolorbox}
    \end{minipage}

    \caption{Study interfaces with key components labeled.}
    \label{fig:study_interfaces}
\end{figure}

For the lay-person study, each participant played 15 games of Blackjack, where the first game was without the model enabled while the other 14 included model predictions. Like with the expert study, the model suggested actions for participants to take. Each game had a maximum between 1 and 7 moves (depending on the cards dealt) with cards drawn from a single deck of cards. We removed betting and added a score which increased or decreased based on the number of wins and losses. Participants could select one of three moves: hit, stand, or surrender.

For each sample labelled / move made participants selected how they used the model. These options were: (1) I was influenced by the \ac{AI}'s suggestion, (2) I was influenced by the concepts the \ac{AI} detected, and (3) I was not influenced by the \ac{AI}. These were designed to capture whether participants selected labels based on the model’s outputs or disregarded them.

Depending on the explanation group, participants were provided with the model's task and concept predictions, saliency maps, and an intervention slider for each concept. Adjusting any intervention slider automatically updated the model's predicted task label. An example of the interfaces are shown in Figure~\ref{fig:study_interfaces}.

At the end of the study, participants completed a closing survey where they were asked to complete questions asking about the model explanations.

\section{Experiment Set-up}

The studies share a number of similarities including interfaces and model capabilities. The studies also apply findings from \cite{Ramaswamy_2023_CVPR} by reducing the number of concepts initially shown to participants by placing them into a scrollable list.

Both studies start with a short demographic survey asking participants for their age, gender, computer science experience and skin disease identification / blackjack experience. Computer science experience and skin disease identification / blackjack experience are recorded using a Likert scale \cite{likert}. Next, participants were briefed on how the model works at a high level and followed a tutorial so they know how to participate in the study and interact with the model. Following this participants compete the study, and finally complete a closing survey.

\subsection{Datasets and Models}

\begin{figure}[t]
     \centering
     \subfloat[Skincon sample with the present concepts ``Plaque'' and ``Patch'']{
         \includegraphics[width=0.3\textwidth]{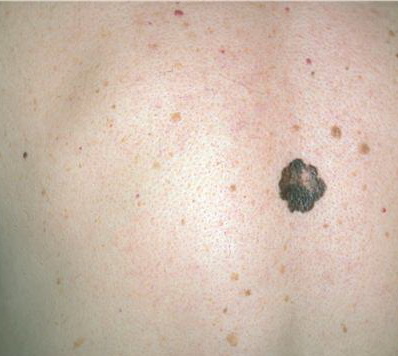}
         \label{fig:example_skincon_sample}
     }
     \hfill
     \subfloat[Standard Blackjack sample]{
         \includegraphics[width=0.3\textwidth]{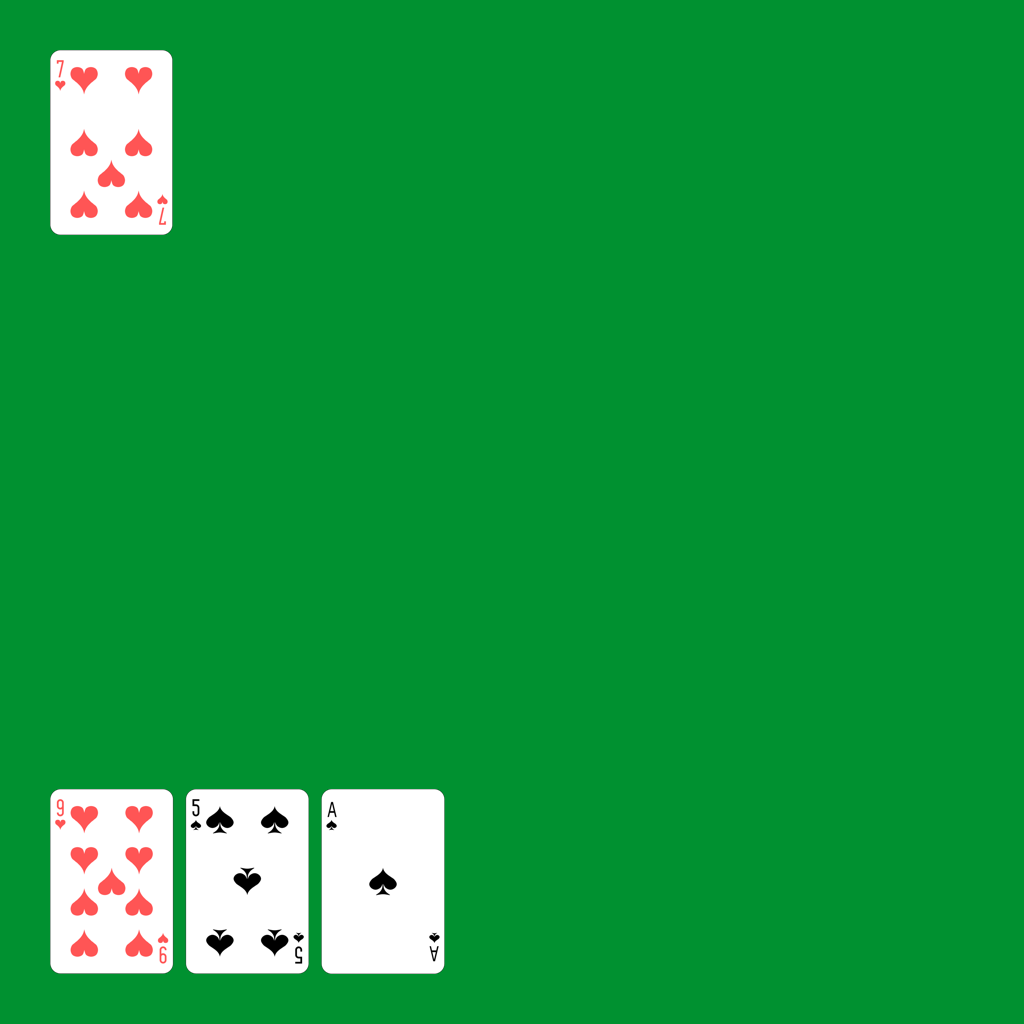}
         \label{fig:example_standard_blackjack}
     }
     \hfill 
     \subfloat[Mixed Blackjack sample]{
         \includegraphics[width=0.3\textwidth]{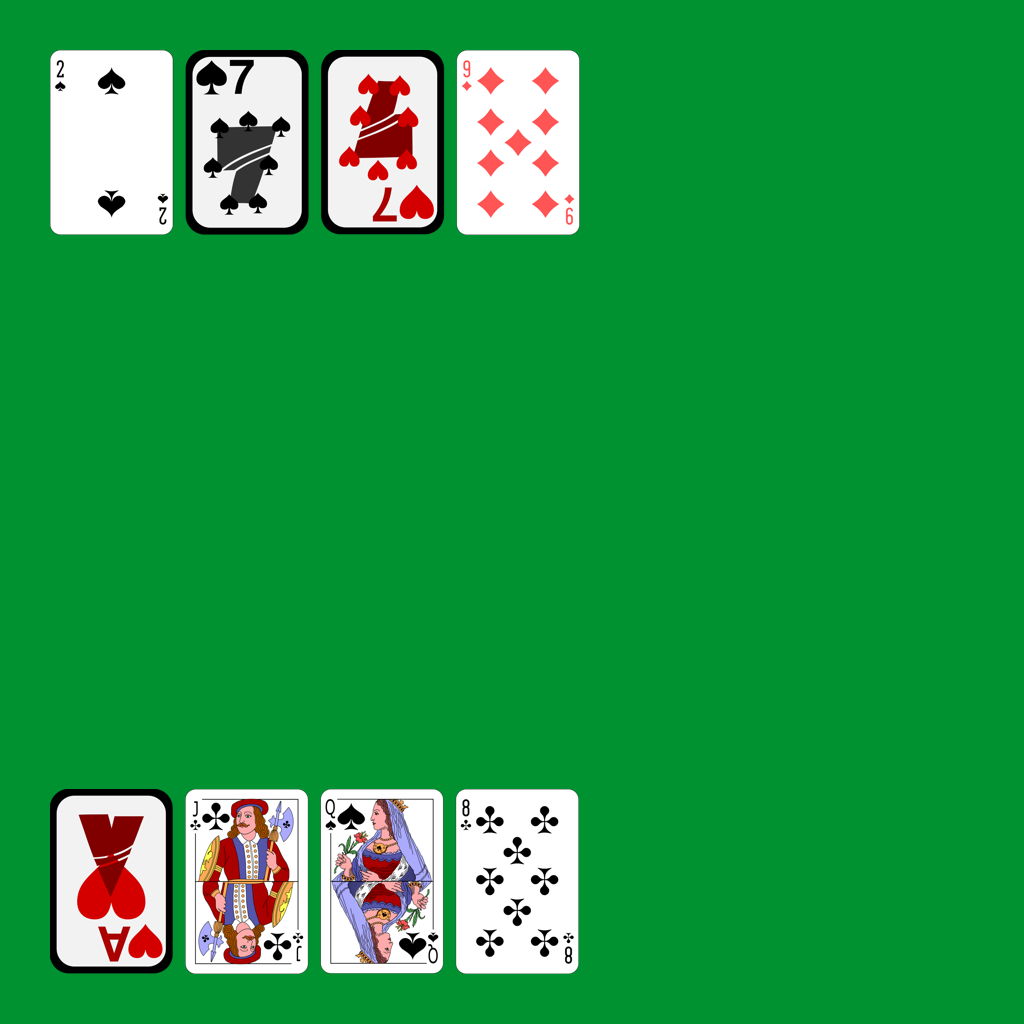}
         \label{fig:example_mixel_blackjack}
     }
     \hfill
     \caption{Example samples from the datasets.}
     \label{fig:example_blackjack}
\end{figure}

\textbf{Expert study}\footnote{Expert study: \url{https://github.com/JackFurby/skin-cbm-study}}: The expert study used a model trained on the Skincon dataset \cite{skincon}, a combination of Fitzpatrick 17k \cite{Fitzpatrick17k} and \ac{DDI} \cite{DDI_dataset}. This is a real-world dataset with 48 clinical concepts, of which we have kept 22 that occur 50 or more times. Concepts were selected by two dermatologists using standard descriptive terms such as ``plaque'' and ``scale''. We have provided an example sample with concept annotation in Figure~\ref{fig:example_skincon_sample}. For task labels, we used the malignant label. We kept the original dataset splits with 10 samples removed from the training and validation splits for the study. In total, we used 2574 train samples, 644 validation samples and 656 test samples.

The Skincon model used a Densenet121 architecture \cite{densenet} for the concept encoder which was initialised with pre-trained weights from ImageNet, and two linear layers with a ReLU activation function for the task predictor which was not pre-trained. The concept encoder was trained to maximise the \ac{AUC} of concept predictions, while the task predictor was trained to minimise task loss. The concept encoder was trained with a learning rate of 0.00053, a \ac{SGD} optimiser and trained for 100 epochs. The task predictor was trained with a learning rate of 0.0593, an Adam optimiser and trained for 100 epochs. The resulting model had a concept accuracy of 91.235\% and a task accuracy of 88.474\%. For the samples included in the study, the model was 70\% accurate.

\textbf{Lay-person study}\footnote{Lay-person study: \url{https://github.com/JackFurby/blackjack-cbm-study}}: For the lay-person study we created the dataset Blackjack\footnote{Blackjack dataset: \url{https://huggingface.co/datasets/JackFurby/blackjack}} which is similar to Playing cards \cite{furby2024constrainconceptbottleneckmodels}. Concepts represent the sum of card values in the player's hand, whether the player has an ``Ace'' card with the value 11, the dealer's first card, and if the dealer has multiple cards. Task labels represent the best move available to the player according to the single deck strategy guide \cite{blackjack-strategy}. These labels are \textit{hit} (player gets another card), \textit{stand} (player ends the game with their current cards), \textit{surrender} (player forfeits their hand for a smaller loss), and \textit{bust} (player's cards sums to over 21).

We created two versions \textit{standard Blackjack}, and \textit{mixed Blackjack}. Standard Blackjack uses one style of playing cards, whereas mixed Blackjack uses a different style for all ``Ace'' and ``Seven'' cards. This allowed us to artificially reduce the accuracy of a model trained on mixed Blackjack if tested on standard Blackjack samples. Each dataset variation has 10,000 samples which are split into training samples and test samples with a 70\%-30\% split respectively. Example samples can be seen in Figures~\ref{fig:example_standard_blackjack} and \ref{fig:example_mixel_blackjack}.

Blackjack models used a VGG-11 architecture with batch normalisation \cite{vgg_models} for the concept encoder and two linear layers with a ReLU activation function for the task predictor. Blackjack models were trained to minimise the concept and task loss. The concept encoder was trained with a learning rate of 0.02 using a \ac{SGD} optimiser. The task predictors had a learning rate of 0.01, used used an Adam optimiser. The models were trained for 200 epochs. The standard Blackjack model achieved a concept accuracy of 99.818\% and a task accuracy of 98.874\%. The mixed Blackjack model achieved a concept accuracy of 96.434\% and a task accuracy of 81.306\%.

\subsection{Evaluation Methodology}

In our studies, we analysed interventions, trust, interpretability, and human-machine performance. To understand when interventions are made we have classified them into two categories: \textit{error correction} and \textit{feature adjustment}. Error correction interventions are concepts that are intervened a maximum of once per sample where the intervened concept value $\bar{c}$ is in the range $0 \leq \bar{c} \leq 0.1$ or $0.9 \leq \bar{c} \leq 1$. Feature adjustments are all other interventions, including concepts that are intervened more than once in a sample, or where the intervened concept value is in the range $0.1 < \bar{c} < 0.9$. Feature adjustment interventions are when the participant is not certain the model has incorrectly predicted the presence of a concept, or where they are inspecting how concepts change task label predictions.

We have also assigned the following labels to understand how the concept value changes with interventions:

\begin{itemize}
    \item Binary change: A concept changes from present to not present, or vice versa.
    \item Changed model task label: An intervention changes the predicted task label.
    \item Magnitude: How much interventions changes concept values by.
    \item Cumulative Change: The total difference between model predicted concept values to the final intervened concept values.
    \item Reversal: Whether final intervened concept values are close to initial concept values.
\end{itemize}

We also tracked interventions over time to evaluate if the rate of interventions increased or decreased. For trust, we evaluated participant and model task label alignment. If the model and human task labels are the same for a large proportion of samples we can argue the human participants trust the model \cite{10.1145/3397481.3450650,10.1145/3287560.3287590}. We also analysed team performance in comparison to ground truth labels from the dataset.

We repeated the test-time intervention metric \cite{concept_bottleneck_models} to measure the change in task accuracy and concepts after interventions. This metric measures the change in model task accuracy when concepts are intervened on. Unlike the original evaluation with \ac{CBM} \cite{concept_bottleneck_models}, our results used human-performed interventions.

At the end of the study participants answered \ac{SCS} \cite{system_causability_scale} questions to measure quality for \ac{AI} explanations. Specifically these questions use a Likert scale and ask participants how they perceived the models ability to answer ``why'' it made task and concept predictions.

\section{Results}

The expert study demographic survey showed participants either agreed or strongly agreed that they can diagnose skin diseases from images. Computer experience was evenly distributed between strongly disagree and agree. In the lay-person study participant demographic showed computer experience increased with blackjack experience.

In the expert study, 120 samples were labelled with 93 interventions performed by 7 out of the 12 participants. Meanwhile, in the lay-person study, 1,456 games of blackjack were played with model move suggestions, and 104 games were played without a model output. 243 interventions were performed by 23 participants out of 52 who had the capability to do so. 65.4\% of interventions were performed on the inaccurate model, and 34.6\% of interventions performed on the accurate model.

\subsection{Intervention Classification}

\begin{table}[t]
    \centering
    \caption{Breakdown of interventions performed in the studies.}
    \begin{tabular}{l|c|c|c|S|S|c|c|c|S|S}
        \hline
        \makecell[tl]{Data subset} & \rotatebox{90}{\shortstack[l]{Total interventions \\(count)}} & \rotatebox{90}{\shortstack[l]{Error correction \\(count)}} & \rotatebox{90}{\shortstack[l]{Feature adjustment \\(count)}} & \rotatebox{90}{\shortstack[l]{Interventions per \\sample (count)}} & \rotatebox{90}{\shortstack[l]{Concept intervened \\per sample (count)}} & \rotatebox{90}{Binary (count)} & \rotatebox{90}{\shortstack[l]{Changed model \\task label (count)}} & \rotatebox{90}{Reversal (count)} & \rotatebox{90}{\shortstack[l]{Mean intervention \\magnitude \\(normalized value)}} & \rotatebox{90}{\shortstack[l]{Mean cumulative \\magnitude \\(normalized value)}} \\
        \hline
        \multicolumn{11}{c}{Expert study} \\
        \hline
        All & 93 & 48 & 45 & 2.91 & 2.50 & 44 & 14 & 11 & 0.48 & 0.5   \\
        \acs{CExp+Int} & 82 & 48 & 34 & 3.04 & 2.78 & 41 & 12 & 6 & 0.49 & 0.52  \\
        \acs{CExp+Int+SMap} & 11 & 0 & 11 & 2.20 & 1 & 3 & 2 & 5 & 0.39 & 0.11   \\
        \hline
        \multicolumn{11}{c}{Lay-person study} \\
        \hline
        All & 243 & 83 & 160 & 4.05 & 2.42 & 152 & 29 & 60 & 0.58 & 0.53   \\
        \makecell[tl]{\acs{Acc}-\acs{CExp+Int}} & 55 & 11 & 44 & 3.93 & 2.71  & 38 & 8 & 7 & 0.59 & 0.57  \\
        \makecell[tl]{\acs{Inacc}-\acs{CExp+Int}} & 73 & 47 & 26 & 3.48 & 2.57 & 41 & 8 & 20 & 0.56 & 0.52  \\
        \makecell[tl]{\acs{Acc}-\acs{CExp+Int+SMap}} & 29 & 0 & 29 & 4.83 & 1.50  & 17 & 5 & 8 & 0.52 & 0.11  \\
        \makecell[tl]{\acs{Inacc}-\acs{CExp+Int+SMap}} & 86 & 25 & 61 & 4.53 & 2.32 & 56 & 8 & 25 & 0.62 & 0.58  \\
        \hline
    \end{tabular}
  \label{tab:interention_classification}
\end{table}

Table~\ref{tab:interention_classification} summarises interventions. Expert study \acs{CExp+Int} participants performed 58.5\% error correction interventions and 41.5\% feature adjustment interventions. 17.6\% of interventions were reversed. 2.78 concepts were intervened per sample with at least one intervention with each intervention changing a concept value by approximately 0.5. Half of all interventions change a concept's presence. 12 interventions changed the model's task prediction, suggesting the model is not sensitive to the concepts participants intervened on.

\acs{CExp+Int+SMap} participants performed fewer interventions with all being feature adjustments and almost all interventions reversed. This indicates all interventions were performed to explore a model's concept sensitivity.

Lay-person study participants demonstrated similar trends with \linebreak\acs{CExp+Int+SMap} participants performing more feature adjustment interventions while \acs{CExp+Int} participants performed a mix of both intervention types. Those with an accurate model primarily performed feature adjustments, aligning with model sensitivity exploration, while those with an inaccurate model split interventions nearly evenly between error correction (47.2\%) and feature adjustments (52.7\%).

\acs{CExp+Int+SMap} participants increased intervention frequency (4.53–4.83 vs. 3.48–3.93 concept intervened), though reduced the number of interventions per sample (1–2 vs. 2–3). Binary interventions were common with inaccurate models, likely due to increased error corrections (47 vs. 11 concepts intervened). High reversal rates for inaccurate models suggest participants required additional interventions to refine their mental model, possibly indicating a misalignment between participants and the model’s decision processes. E.g. if the participant plays with a different strategy.

\subsection{Human-machine Task Alignment}

We measured the alignment between participant-selected task labels and model predicted task labels to determine if participants trusted the model or not. To further reinforce if alignment is helping the human-machine team, we also compare team accuracy to determine if trust is justified.

\begin{table}[t]
    \centering
    \caption{Expert study human-machine task alignment.}
    \begin{tabular}{l|c|c|c|c}
        \hline
        \makecell[tl]{Data subset} & \rotatebox{90}{Overall (\%)} & \rotatebox{90}{\makecell[tl]{Initial model task \\prediction (\%)}} & \rotatebox{90}{\makecell[tl]{Intermediate \\model task \\prediction (\%)}} & \rotatebox{90}{\makecell[tl]{Final model task \\prediction (\%)}} \\
        \hline

        All & 80.8 (±3.6) & 77.5 (±3.8) & 77.8 (±8.2) & 65.6 (±8.5) \\
        \acs{CExp+Int} & \textbf{81.7 (±5.0)} & 76.7 (±5.5) & 81.8 (±8.4) & 70.4 (±9.0) \\
        \acs{CExp+Int+SMap} & 80.0 (±5.2) & 78.3 (±5.4) & 60.0 (±24.5) & 40.0 (±24.5) \\
        \acs{NoInt} & \textbf{81.8 (±4.1)} & 81.8 (±4.1) & - & - \\
        \acs{WithInt} & 78.1 (±7.4) & 65.6 (±8.5) & 77.8 (±8.2) & 65.6 (±8.5) \\
        \acs{CExp+Int}-\acs{NoInt} & \textbf{81.8 (±6.8)} & 81.8 (±6.8) & - & - \\
        \acs{CExp+Int}-\acs{WithInt} & 81.5 (±7.6) & 70.4 (±9.0) & 81.8 (±8.4) & 70.4 (±9.0) \\
        \makecell[tl]{\acs{CExp+Int+SMap}-\acs{NoInt}} & \textbf{81.8 (±5.2)} & 81.8 (±5.2) & - & - \\
        \makecell[tl]{\acs{CExp+Int+SMap}-\acs{WithInt}} & 60.0 (±24.5) & 40.0 (±24.5) & 60.0 (±24.5) & 40.0 (±24.5) \\
        \hline
    \end{tabular}
    \label{tab:skincon_human_ai_alignment}
\end{table}

Table~\ref{tab:skincon_human_ai_alignment} shows expert study alignment. As there is no guarantee the final model task prediction is the model prediction that a participant aligns with, the overall alignment includes agreement with initial and post-intervention task labels. Alignment for this ranges from 60\% to 81.8\%. Initial alignment reflects the model’s original label and is consistently higher for participants without interventions, final alignment only includes the model's last task label after interventions, and intermediate alignment includes model labels between the initial model task prediction and final task prediction.

A slight decline in alignment is observed with \acs{WithInt} participants (78.1\% vs. 81.8\%) and \acs{CExp+Int+SMap} participants (80\% vs. 81.7\%). These findings suggest interventions influence participants’ labelling decisions, reducing agreement with the model. In fact, initial alignment with interventions (65.6\%) is closer to the model’s actual accuracy (70\%). This indicates interventions help participants calibrate trust to align with the model's true accuracy. Meanwhile, \acs{NoInt} participants appear to over-trust the model. A one-tailed t-test reveals statistical significance with a p-value of 0.03, below the 0.05 threshold, confirming that the absence of interventions led to increased alignment in this study. However, the difference in alignment between \acs{CExp+Int} and \acs{CExp+Int+SMap} participants was not statistically significant (p-value of 0.59), suggesting that saliency maps had no meaningful impact on alignment. However, a larger study may confirm otherwise.

\begin{table}[ht!]
    \centering
    \caption{Lay-person study human-machine task alignment.}
    \begin{tabular}{l|c|c|c|c}
        \hline
        \makecell[tl]{Data subset} & \rotatebox{90}{Overall (\%)} & \rotatebox{90}{\makecell[tl]{Initial model task \\prediction (\%)}} & \rotatebox{90}{\makecell[tl]{Intermediate \\model task \\prediction (\%)}} & \rotatebox{90}{\makecell[tl]{Final model task \\prediction (\%)}} \\
        \hline

        All & 77.3 (±0.8) & 77.1 (±0.8) & 81.5 (±5.3) & 83.3 (±4.9) \\
        \acs{NoInt} & 77.1 (±0.8) & 77.1 (±0.8) & - & - \\
        \acs{WithInt} & \textbf{86.7 (±4.4)} & 76.7 (±5.5) & 81.5 (±5.3) & 83.3 (±4.9) \\
        \acs{Acc} & \textbf{80.1 (±1.1)} & 80.0 (±1.1) & 93.8 (±6.2) & 90.0 (±6.9) \\
        \acs{Inacc} & 74.6 (±1.2) & 74.3 (±1.2) & 76.3 (±7.0) & 80.0 (±6.4) \\
        \acs{Acc}-\acs{NoExp} & \textbf{79.8 (±2.2)} & 79.8 (±2.2) & - & - \\
        \acs{Inacc}-\acs{NoExp} & 70.4 (±2.6) & 70.4 (±2.6) & - & - \\
        \acs{Acc}-\acs{CExp} & \textbf{84.5 (±2.0)} & 84.5 (±2.0) & - & - \\
        \acs{Inacc}-\acs{CExp} & 73.9 (±2.5) & 73.9 (±2.5) & - & - \\
        \acs{Acc}-\acs{CExp+Int}-\acs{NoInt} & 78.8 (±2.4) & 78.8 (±2.4) & - & - \\
        \makecell[tl]{\acs{Acc}-\acs{CExp+Int}-\acs{WithInt}} & \textbf{92.9 (±7.1)} & 78.6 (±11.4) & 90.0 (±10.0) & 92.9 (±7.1) \\
        \acs{Inacc}-\acs{CExp+Int}-\acs{NoInt} & 74.8 (±2.5) & 74.8 (±2.5) & - & - \\
        \makecell[tl]{\acs{Inacc}-\acs{CExp+Int}-\acs{WithInt}} & \textbf{76.2 (±9.5)} & 66.7 (±10.5) & 73.7 (±10.4) & 71.4 (±10.1) \\
        \makecell[tl]{\acs{Acc}-\acs{CExp+Int+SMap}-\acs{NoInt}} & 76.0 (±2.5) & 76.0 (±2.5) & - & - \\
        \makecell[tl]{\acs{Acc}-\acs{CExp+Int+SMap}-\acs{WithInt}} & \textbf{100 (±0.0)} & 100 (±0.0) & 100 (±0.0) & 83.3 (±16.7) \\
        \makecell[tl]{\acs{Inacc}-\acs{CExp+Int+SMap}-\acs{NoInt}} & 78.5 (±2.4) & 78.5 (±2.4) & - & - \\
        \makecell[tl]{\acs{Inacc}-\acs{CExp+Int+SMap}\\-\acs{WithInt}} & \textbf{89.5 (±7.2)} & 78.9 (±9.6) & 78.9 (±9.6) & 89.5 (±7.2) \\
        \hline
    \end{tabular}
    \label{tab:blackjack_human_ai_alignment}
\end{table}

Alignment for the lay-person study is shown in Table~\ref{tab:blackjack_human_ai_alignment} averages to 77.3\%, which is lower than the model task accuracy (99.8\% for the accurate model and 96.4\% for the inaccurate model). However, participants may have playing strategies misaligned with the training data. Alignment in this study differs from the expert study as interventions increase alignment (86.7\% vs. 77.1\%). Further, participants using the accurate model consistently had a higher alignment than the inaccurate model.

Across all participant groups, interventions improve human-machine alignment. Alignment also increases from the initial model prediction to the final model prediction for most participant subsets. A one-tailed t-test supports this by rejecting the null hypothesis (no alignment increase) and accepting the alternative (interventions improve alignment), with p-values of 0.041 for all participants and 0.04 for those capable of performing interventions, both below the 0.05 significance threshold. Additionally, comparing \acs{CExp} participants to \acs{NoExp} participants resulted in a p-value of 0.036, showing concept explanations also result in a higher human-machine alignment. These results show both interventions and concept explanations increase human-machine task alignment. 

\begin{table}[t]
    \centering
    \caption{Expert study human-machine task accuracy.}
    \begin{tabular}{l|c|c|c}
        \hline
        Data Subset & \rotatebox{90}{\makecell[tl]{Overall \\Accuracy (\%)}} & \rotatebox{90}{\makecell[tl]{Malignant \\Melanoma(\%)}} & \rotatebox{90}{\makecell[tl]{Seborrheic \\Keratosis (\%)}} \\
        \hline

        All & 78.3 (±2.4) & 88.3 (±4.6) & 68.3 (±3.9) \\
        \acs{CExp+Int} & 75.0 (±3.4) & 83.3 (±8.0) & 66.7 (±4.2) \\
        \acs{CExp+Int+SMap} & \textbf{81.7 (±3.1)} & 93.3 (±4.2) & 70.0 (±6.8) \\
        \acs{NoInt} & 78.0 (±3.7) & 88.0 (±8.0) & 68.0 (±8.0) \\
        \acs{WithInt} & \textbf{78.6 (±3.4)} & 88.6 (±5.9) & 68.6 (±4.0) \\
        \acs{CExp+Int}-\acs{NoInt} & 75.0 (±5.0) & 80.0 (±20.0) & 70.0 (±10.0) \\
        \acs{CExp+Int}-\acs{WithInt} & 75.0 (±5.0) & 85.0 (±9.6) & 65.0 (±5.0) \\
        \acs{CExp+Int+SMap}-\acs{NoInt} & 80.0 (±5.8) & 93.3 (±6.7) & 66.7 (±13.3) \\
        \acs{CExp+Int+SMap}-\acs{WithInt} & \textbf{83.3 (±3.3)} & 93.3 (±6.7) & 73.3 (±6.7) \\
        \hline
    \end{tabular}
    
    \label{tab:skincon_human_ai_accuracy_table}
\end{table}

If appropriate trust is given to the model, we should expect the human-machine accuracy to be higher than the model alone. Table~\ref{tab:skincon_human_ai_accuracy_table} shows human-machine accuracy for the expert study. Accuracy is averaged by participants, assuming that they build a mental model of the model over time. Even if an individual prediction is ignored, they may still influence participants. For instance, participants may recognise when the model is incorrect without the need to perform interventions.

In the expert study \acs{CExp+Int} participants achieved an accuracy of 75\% and \acs{CExp+Int+SMap} participants achieved an accuracy of 81.7\%, indicating that the additional information provided by saliency maps aids decision-making. \acs{WithInt} Participants had a slightly higher accuracy than \acs{NoInt} participants (78.6\% vs. 78\%), suggesting interventions either match or slightly enhance participant performance. However, the expert study lacks the sample size to show statistical significance. A one-tailed t-test resulted in a p-value of 0.09 for accuracy being higher if participants had saliency maps, and a p-value of 0.46 if participants performed interventions.

\begin{table}[t]
    \centering
    \caption{lay-person study human-machine task accuracy averaged by participant.}
    \begin{tabular}{l|c}
        \hline
        Data Subset & Accuracy (\%)  \\
        \hline

\ac{AI} disabled & 74.4 (±3.9) \\
All & 83.6 (±0.9) \\
\acs{WithInt} & 83.3 (±1.1) \\
\acs{NoInt} & \textbf{84.7 (±1.8)} \\
\acs{Acc} & \textbf{84.6 (±2.7)} \\
\acs{Inacc} & 78.1 (±3.2) \\
\acs{Acc}-\acs{NoExp} & \textbf{84.6 (±2.7)} \\
\acs{Inacc}-\acs{NoExp} & 78.1 (±3.2) \\
\acs{Acc}-\acs{CExp} & \textbf{91.0 (±2.4)} \\
\acs{Inacc}-\acs{CExp} & 81.4 (±1.6) \\
\acs{Acc}-\acs{CExp+Int}-\acs{NoInt} & \textbf{86.6 (±3.3)} \\
\acs{Acc}-\acs{CExp+Int}-\acs{WithInt} & 83.8 (±2.9) \\
\acs{Inacc}-\acs{CExp+Int}-\acs{NoInt} & 75.7 (±4.6) \\
\acs{Inacc}-\acs{CExp+Int}-\acs{WithInt} & \textbf{84.3 (±1.8)} \\
\acs{Acc}-\acs{CExp+Int+SMap}-\acs{NoInt} & \textbf{83.4 (±2.4)} \\
\acs{Acc}-\acs{CExp+Int+SMap}-\acs{WithInt} & \textbf{83.4 (±6.4)} \\
\acs{Inacc}-\acs{CExp+Int+SMap}-\acs{NoInt} & 83.5 (±2.9) \\
\acs{Inacc}-\acs{CExp+Int+SMap}-\acs{WithInt} & \textbf{86.6 (±3.6)} \\
\hline
    \end{tabular}
    \label{tab:blackjack_human_ai_accuracy_table}
\end{table}

In the lay-person study (Table~\ref{tab:blackjack_human_ai_accuracy_table}) the largest increase in task accuracy came from participants using the accurate model compared to the inaccurate model. Interventions only increased human-machine accuracy for participants using the inaccurate model. This suggests participants are over-trusting the model, expectedly considering interventions increased alignment as just discussed in Table~\ref{tab:blackjack_human_ai_alignment}. Despite this, the lowest task accuracy was achieved by  \acs{Inacc}-\acs{CExp} and \acs{Inacc}-\acs{NoExp} participants and participants with the \ac{AI} disabled. Therefore interventions are still showing signs of increasing human-machine accuracy.

A one-tailed t-test resulted in a p-value of 0.042 showed \acs{CExp} improved task accuracy compared to \acs{NoExp}. In contrast, participants who performed interventions did not achieve a statistically significant improvement in task accuracy compared to those who did not, achieving a p-value of 0.27 when compared to all \acs{NoExp} and \acs{CExp} participant groups and 0.195 when compared to participants who did not perform interventions but had the capability to do so. While we observed a trend of higher accuracy among participants using interventions, we cannot conclude that interventions directly improve task accuracy.

Overall, our findings indicate that concepts are beneficial for improving human-machine alignment and human-machine task accuracy. Interventions are beneficial for increasing human-machine task alignment but do not result in a statistically significant increase in task accuracy and can result in over-trust.

\subsection{Interventions Over Time}

\begin{figure}[ht!]
    \centering
    \subfloat[Expert study]{
        \includegraphics[width=0.3\textwidth]{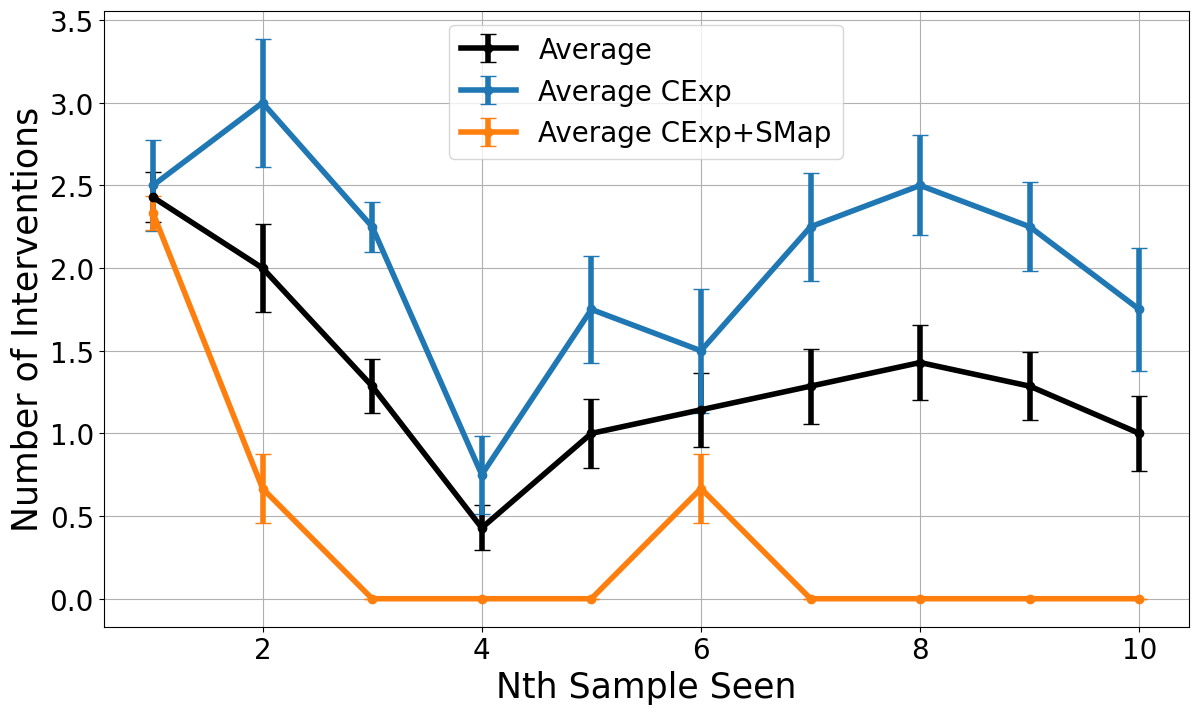}
        \label{fig:skincon_intervention_time}
    }
    \hfill
    \subfloat[Lay-person study correctly predicted concepts]{
        \includegraphics[width=0.3\textwidth]{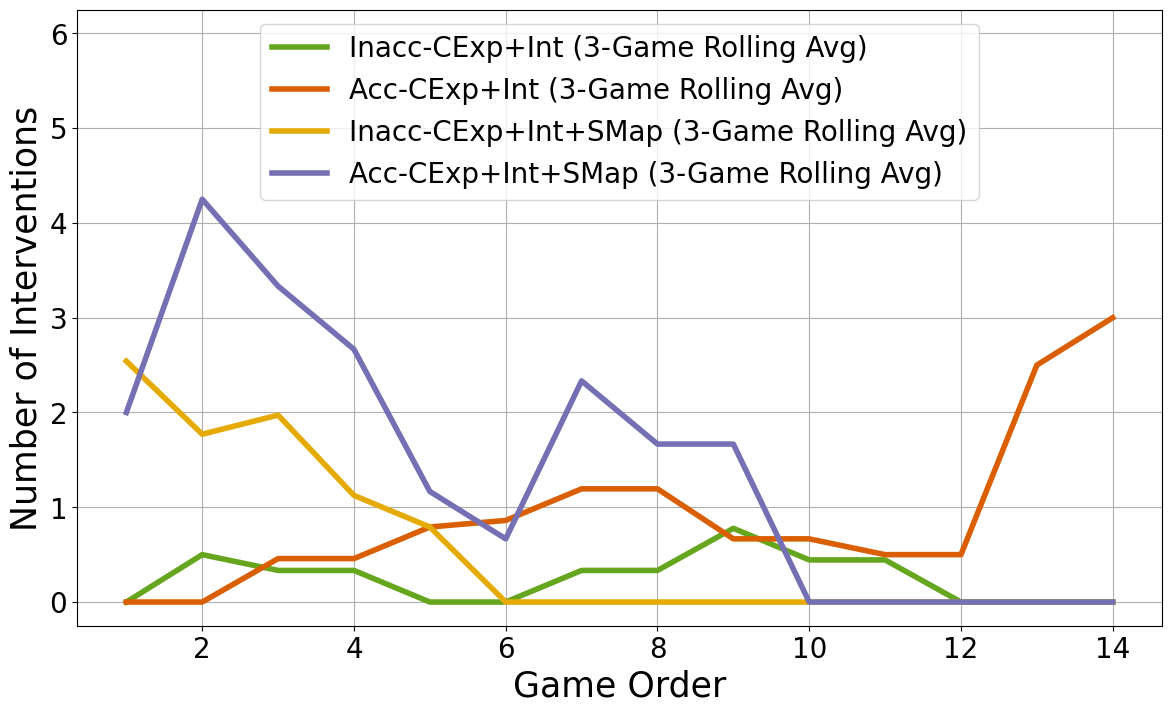}
        \label{fig:blackjack_interventions_accurate_concepts}
    }
    \hfill
    \subfloat[Lay-person study incorrectly predicted concepts]{
        \includegraphics[width=0.3\textwidth]{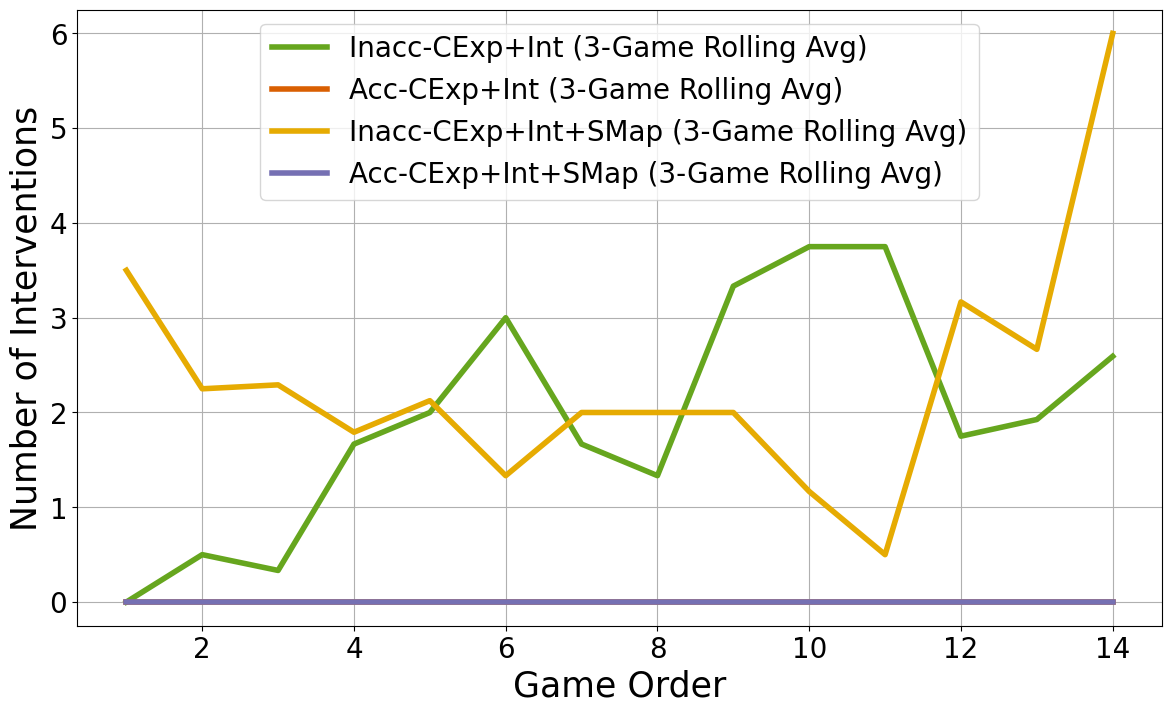}
        \label{fig:blackjack_interventions_inaccurate_concepts}
    }
    \caption{Interventions performed declined over time except for incorrectly predicted concepts in the lay-person study where the number of interventions performed remains constant.}
    \label{fig:blackjack_interventions_per_game}
\end{figure}

We may expect the number of interventions to decrease over time as participants learn about a model's sensitivity to concepts. We found this is the case in both the expert study (Figure~\ref{fig:skincon_intervention_time}) where we show the average number of interventions per sample with the standard error, and the layperson study when the model correctly predicts concepts (Figure~\ref{fig:blackjack_interventions_accurate_concepts}) where we show the decline with a rolling average. An exception to the decline is seen with \acs{Acc}-\acs{CExp+Int} participants in the lay-person study where there is a spike of interventions performed on game 13 with an average of 7 and a standard error of ±2. As this occurs once, this spike is not representative of all participants. 

In the expert study, \acs{CExp+Int+SMap} participants see a sharp decline in interventions, while \acs{CExp+Int}  participants see an initial decline which recovers for later samples. This suggests saliency map explanations provide additional insights into the model's concept predictions. \acs{CExp+Int}  participants appear to be incentivised to use interventions as a means of understanding the model's decision-making process.

For incorrect concept predictions in the lay-person study, participants consistently performed around 2 interventions per sample. These results demonstrate participants identify concepts that need to be intervened on while ignoring the concepts that are correctly predicted by the models.

Overall, these results show that participants initially explore the model's capabilities and sensitivity to concept values before developing a mental model and reducing the number of interventions performed to where it is required.

\subsection{Test-time Intervention and Concept Accuracy}

The authors of \acp{CBM} showed results using the metric test-time intervention where interventions updating concept predictions with ground truth concept values improved model task performance \cite{concept_bottleneck_models}. However, it remains unknown how interventions improve model task performance when interventions are made by humans.

In the related work section, we discussed \cite{Barker2023SelectiveCM} where they found \acp{CBM} may not apply the same weight to concepts for task labels as humans would. Therefore we hypothesise interventions performed by humans may not see an improvement in task accuracy which would show a misalignment between humans and the model's sensitivity to concepts.

\begin{figure}[ht!]
    \centering
    \subfloat[Task accuracy]{
        \includegraphics[width=0.3\textwidth]{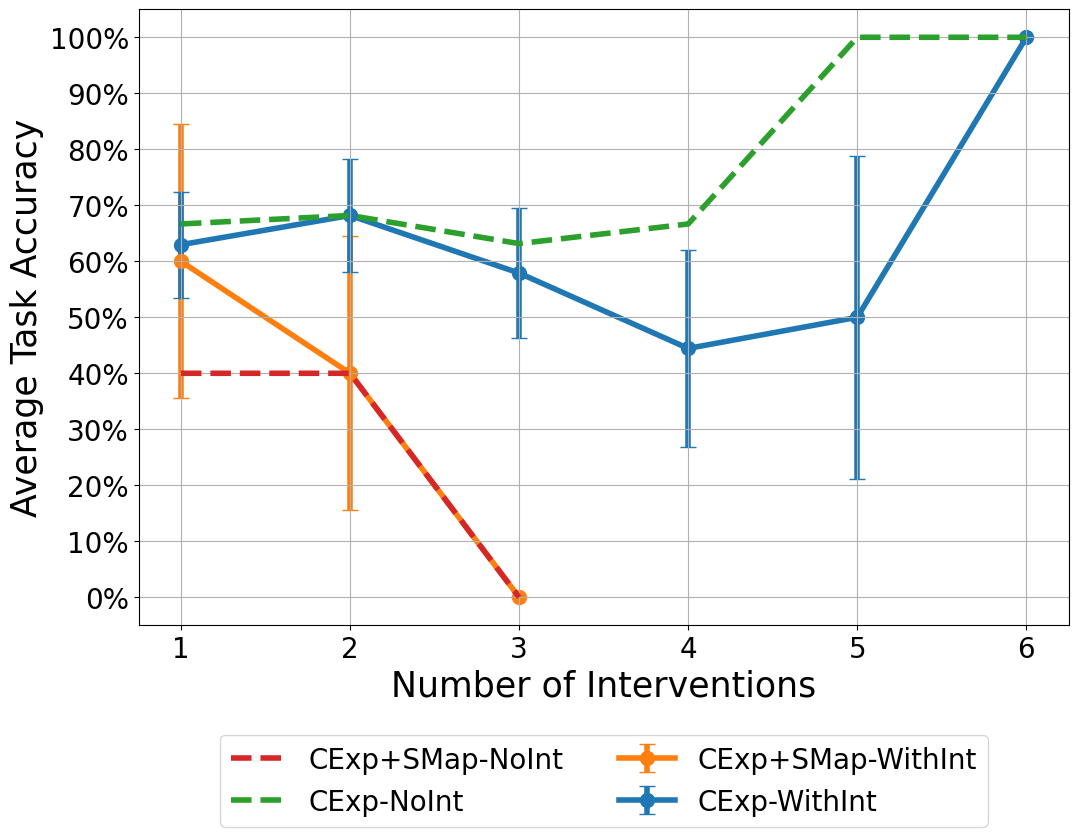}
        \label{fig:skincon_tti_task}
    }
    \hfill
    \subfloat[Concept precision]{
        \includegraphics[width=0.3\textwidth]{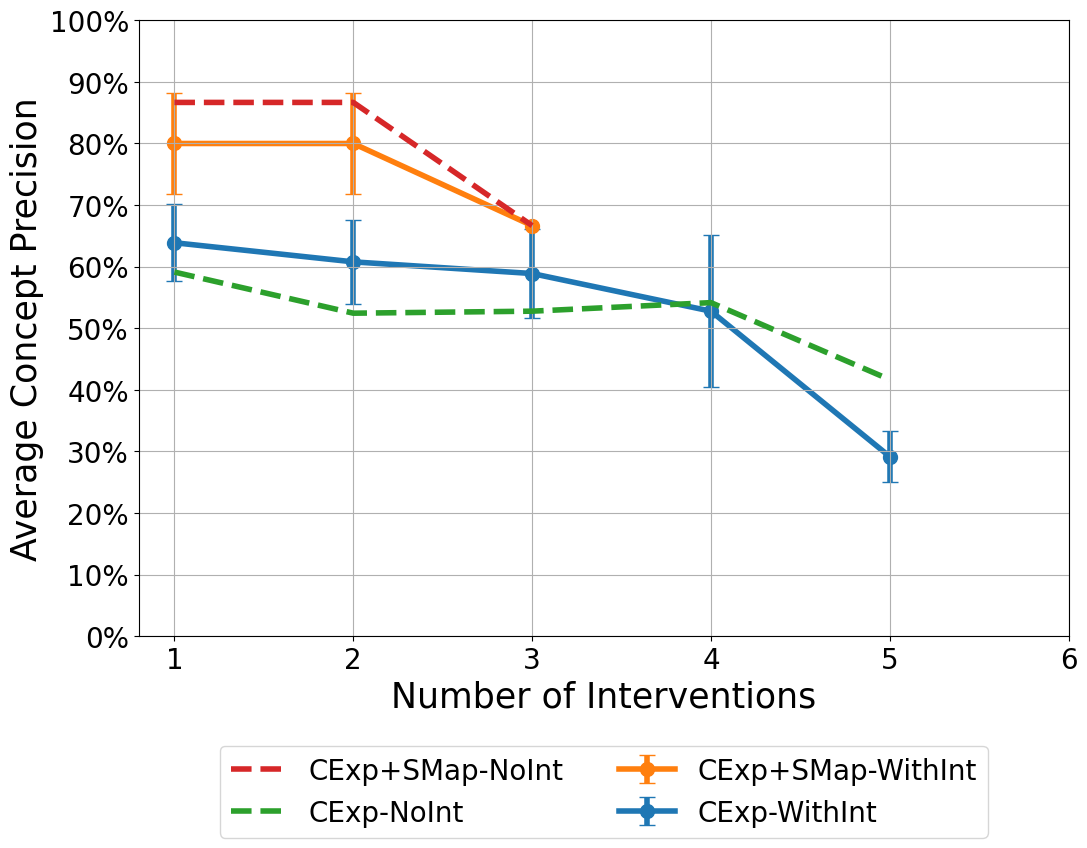}
        \label{fig:skincon_concept_precision}
    }
    \hfill
    \subfloat[Concept recall]{
        \includegraphics[width=0.3\textwidth]{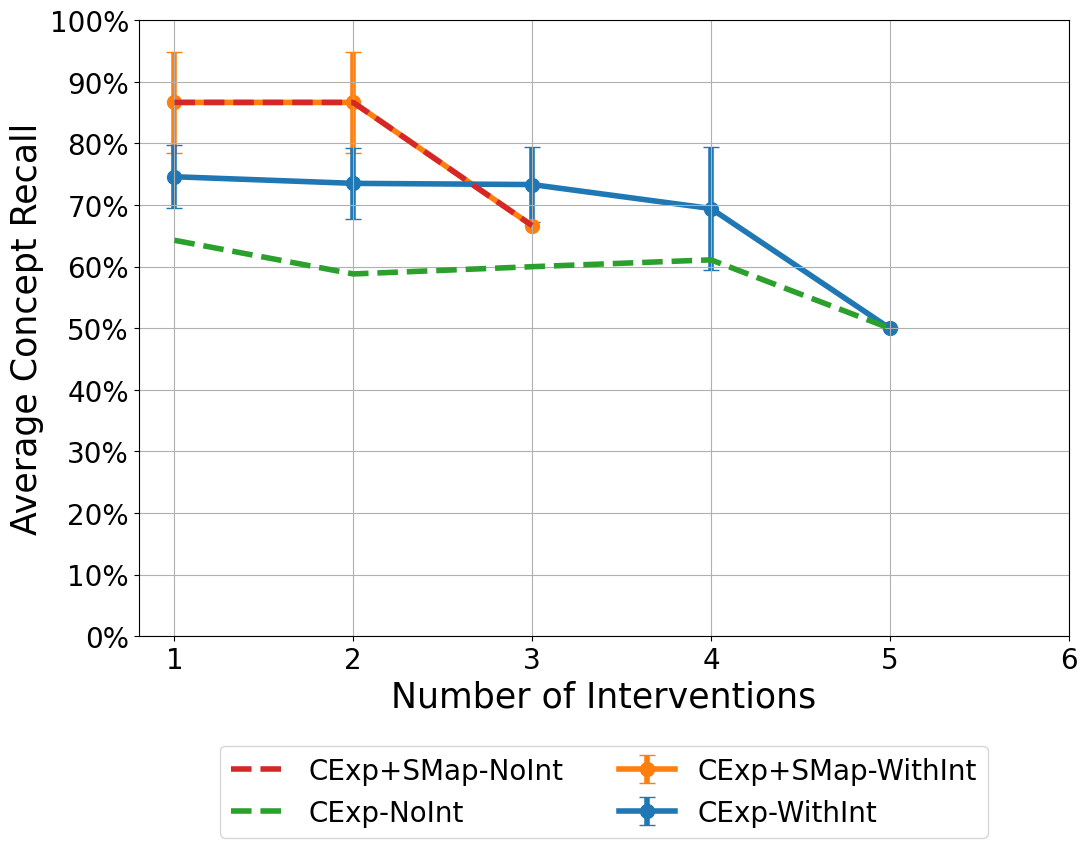}
        \label{fig:skincon_concept_recall}
    }
    \hfill
    \subfloat[Accurate model task]{
        \includegraphics[width=0.3\textwidth]{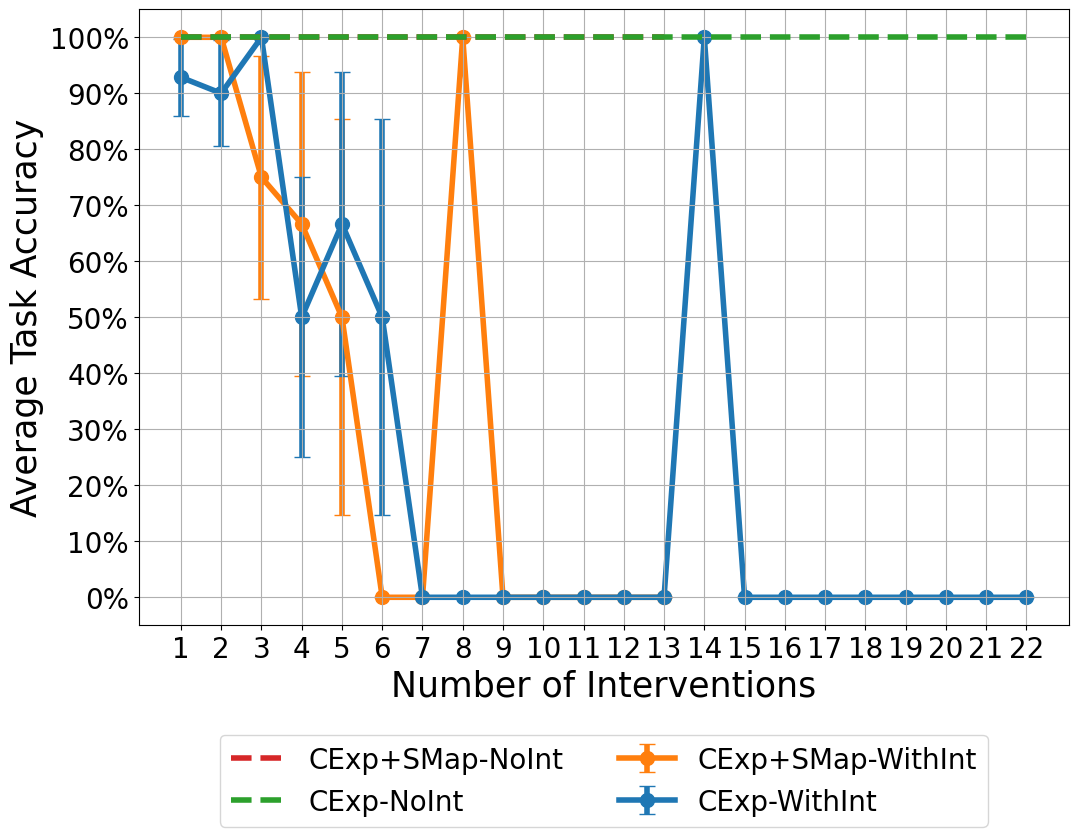}
        \label{fig:blackjack_tti_task_accurate}
    }
    \hfill
    \subfloat[Accurate model concept precision]{
        \includegraphics[width=0.3\textwidth]{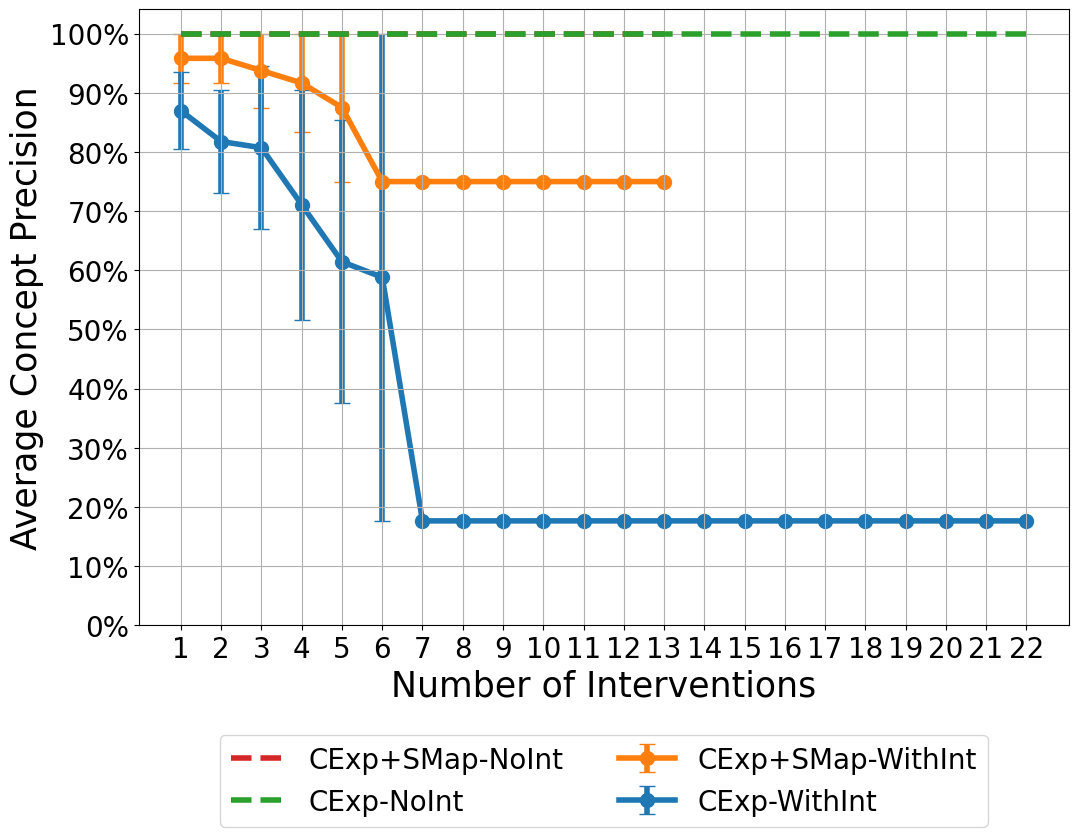}
        \label{fig:blackjack_tti_precision_accurate}
    }
    \hfill
    \subfloat[Accurate model concept recall]{
        \includegraphics[width=0.3\textwidth]{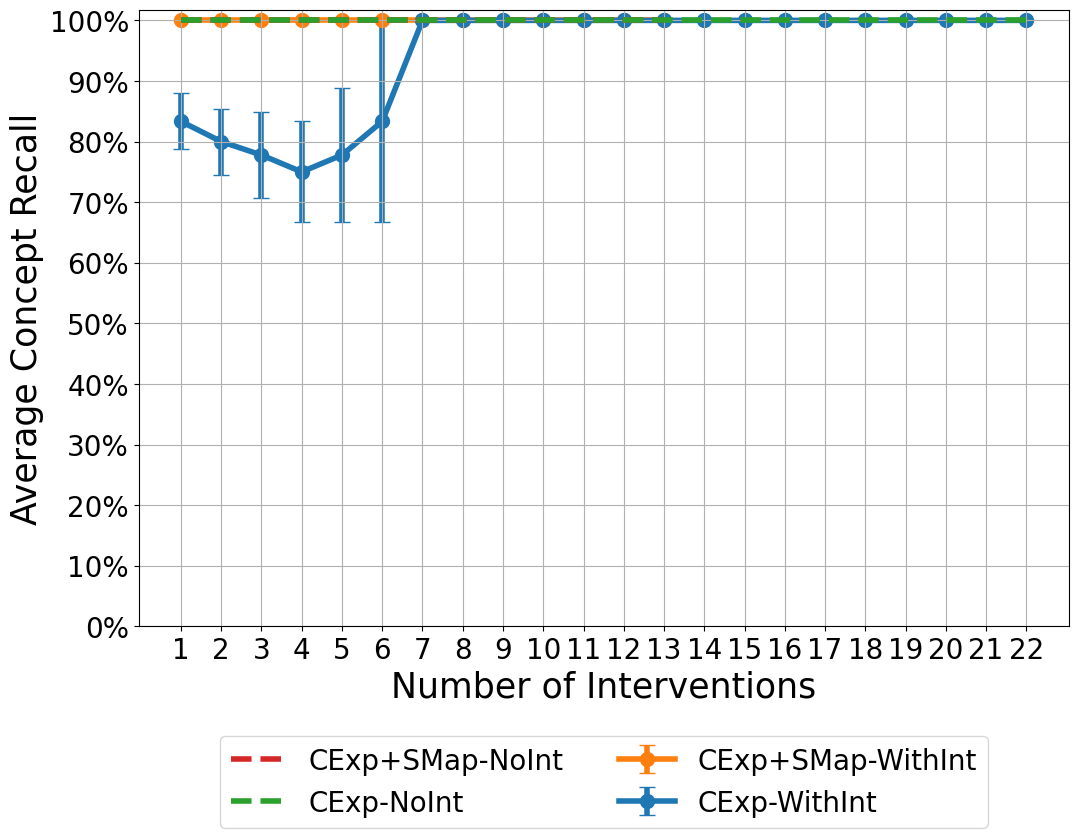}
        \label{fig:blackjack_recall_accurate}
    }
    \hfill
    \subfloat[Inaccurate model task]{
        \includegraphics[width=0.3\textwidth]{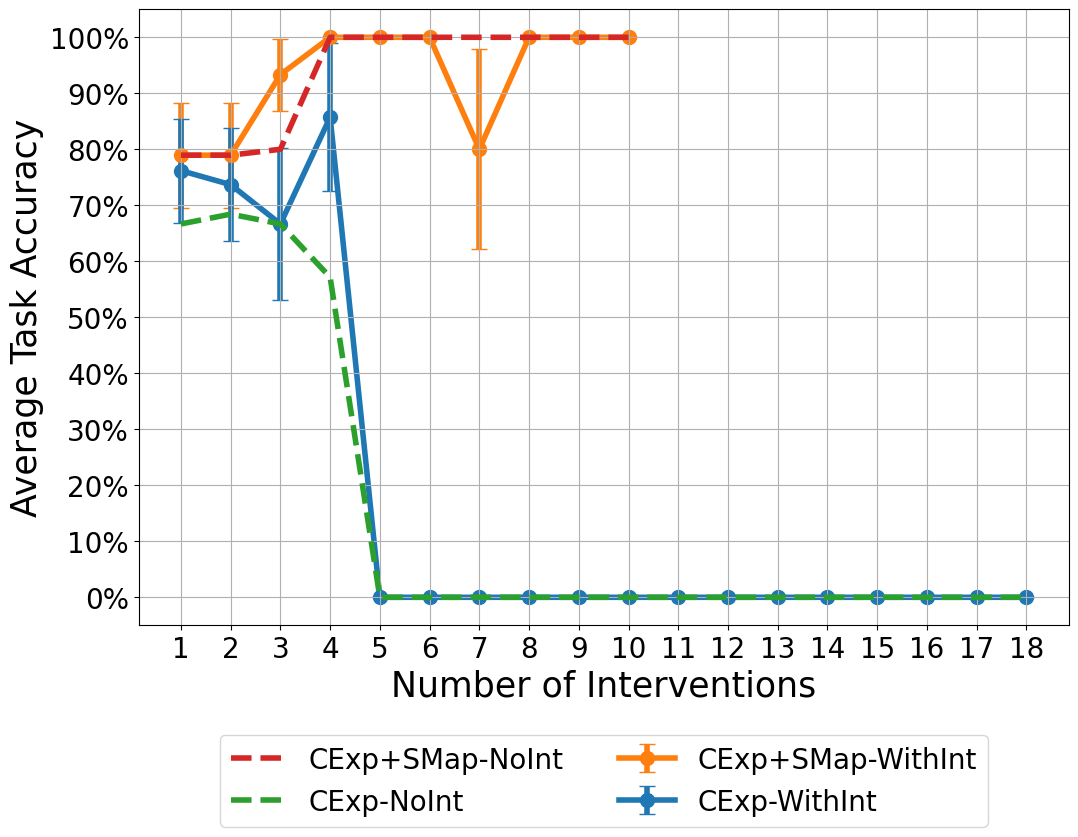}
        \label{fig:blackjack_tti_task_inaccurate}
    }
    \hfill
    \subfloat[Inaccurate model concept precision]{
        \includegraphics[width=0.3\textwidth]{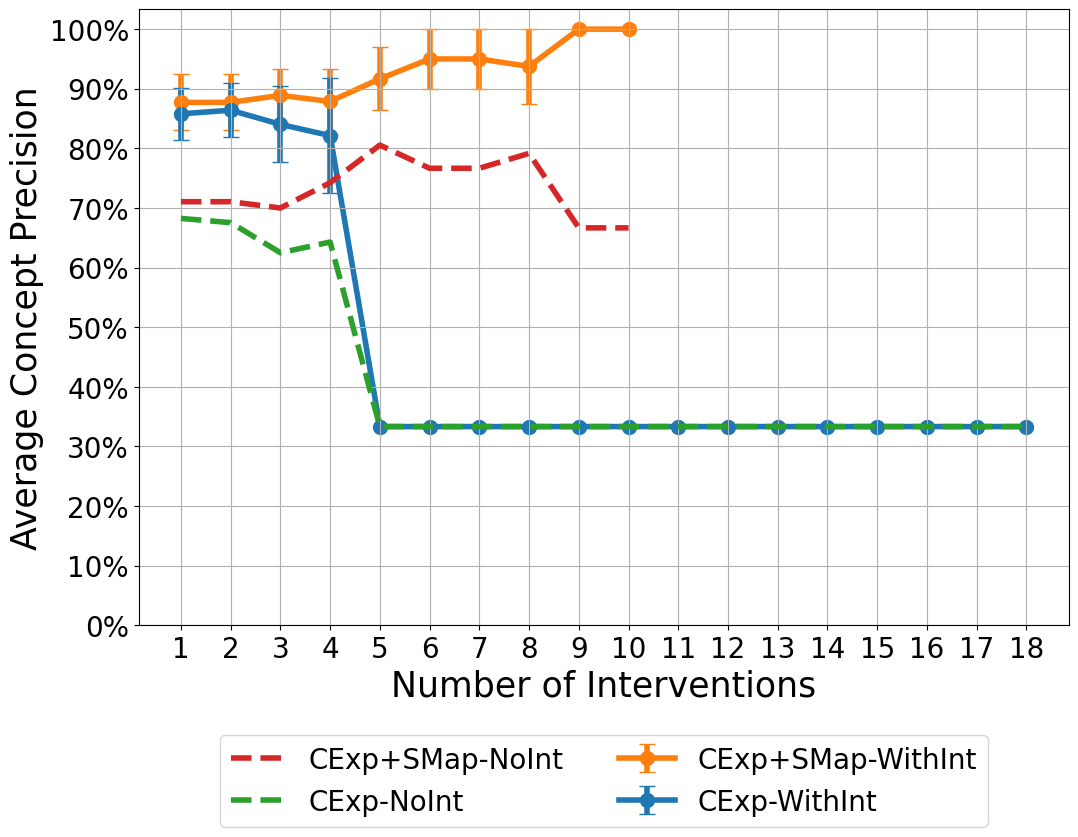}
        \label{fig:blackjack_tti_precision_inaccurate}
    }
    \hfill
    \subfloat[Inaccurate model concept recall]{
        \includegraphics[width=0.3\textwidth]{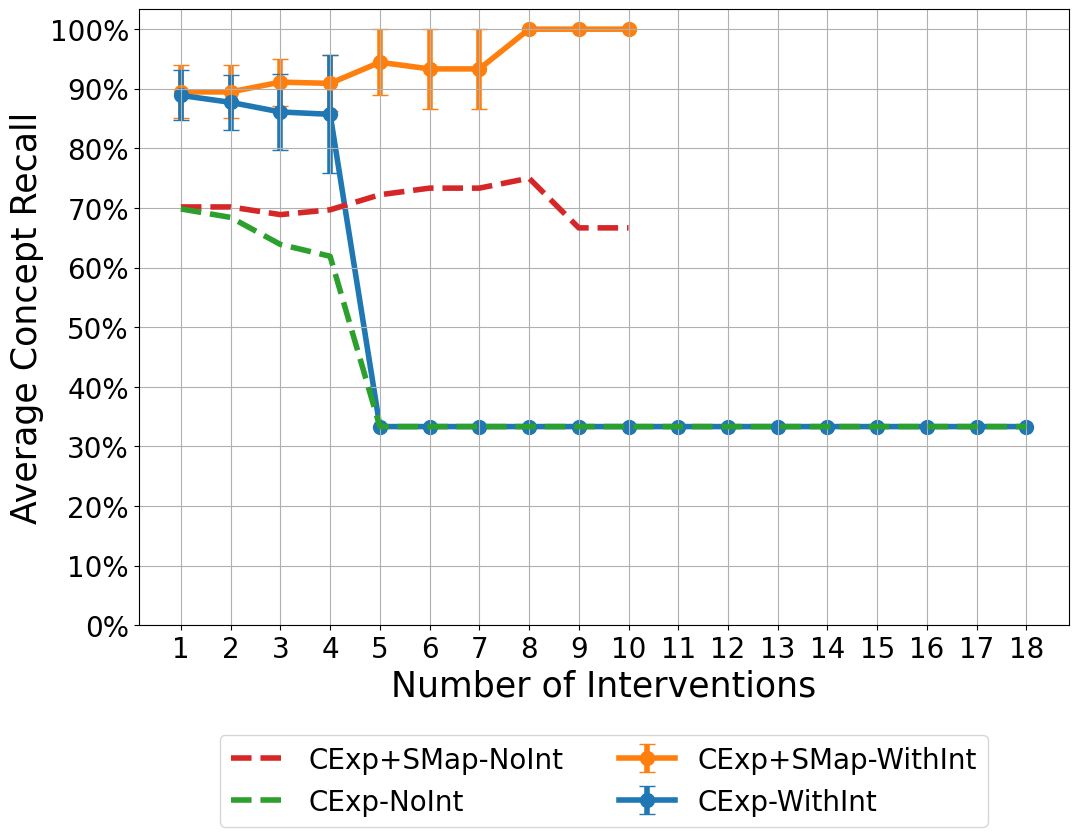}
        \label{fig:blackjack_recall_inaccurate}
    }
    \caption{Interventions decrease model task accuracy in the expert study and the layperson study accurate model while increasing model task accuracy with the lay-person study inaccurate model. Concept precision and recall increase with interventions in most cases.}
    \label{fig:tti}
\end{figure}

Test-time intervention results are shown in Figure~\ref{fig:skincon_tti_task} for the expert study and Figures~\ref{fig:blackjack_tti_task_accurate} and \ref{fig:blackjack_tti_task_inaccurate} for the lay-person study. Task accuracy is averaged by participant and explanation groups. We only included the same samples between with interventions and without. For example, if participants intervened on concepts for samples 1 and 3 but not 2, we only work out the task accuracy for samples 1 and 3. As participants performed 2 - 3 interventions on average in each study conclusions for concept and task accuracy past these intervention counts cannot always be made.

Task accuracy, for the most part, does not improve with interventions compared to no interventions. In the expert study between 1 - 3 interventions task accuracy is close to matching the accuracy of the model with no interventions, and outperforms the model with 1 intervention for \acs{CExp+Int+SMap} participants. In the lay-person study task accuracy initially declines slightly for the accurate model before sharply declining, although with large error bars. The exception to declining task accuracy is observed for the lay-person study inaccurate model where interventions increase or match the model's initial task accuracy.

In addition to task accuracy, we have also measured the change in concept accuracy. Figure~\ref{fig:skincon_concept_precision} and Figure~\ref{fig:skincon_concept_recall} shows the precision and recall for the expert study, Figure~\ref{fig:blackjack_tti_precision_accurate} and Figure~\ref{fig:blackjack_recall_accurate} for the lay-person study accurate model, and Figure~\ref{fig:blackjack_tti_precision_inaccurate} and Figure~\ref{fig:blackjack_recall_inaccurate} for the inaccurate model. Notably, in the expert study and lay-person study with the inaccurate model, interventions lead to an increase or matching the precision and recall. In the lay-person study with the accurate model, precision is lower than the model concept prediction, while recall initially declines before rising the match the model. Most of these results also have no overlapping error bars between the model-predicated concepts and intervened concepts.

\begin{table}[ht]
    \centering
    \caption{Likert Scores for \ac{SCS} questions.}
    \begin{tabular}{l|c|c|c|c|c|c|c}
    \hline
    Question & \rotatebox{90}{All} & \rotatebox{90}{\acs{CExp+Int}} & \rotatebox{90}{\acs{CExp+Int+SMap}} & \rotatebox{90}{\acs{WithInt}} & \rotatebox{90}{\acs{NoInt}} & \rotatebox{90}{\makecell[tl]{Skin Experience \\Agree}} & \rotatebox{90}{\makecell[tl]{Skin Experience \\Strongly Agree}} \\ \hline
    Factors in data& 3.09 & 3.00 & 3.20 & 3.00 & 3.20 & 2.80 & \textbf{3.33} \\
    Understood& 3.73 & 3.33 & 4.20 & 3.67 & 3.80 & 3.60 & \textbf{3.83} \\
    Change detail level& 3.18 & \textbf{3.67} & 2.60 & 3.50 & 2.80 & 2.80 & 3.50 \\
    Need support& 3.64 & 3.33 & \textbf{4.00} & 3.50 & 3.80 & \textbf{4.00} & 3.33 \\
    Understanding causality& 3.00 & 3.17 & 2.80 & \textbf{3.33} & 2.60 & 2.80 & 3.17 \\
    Use with knowledge& 3.45 & 3.50 & 3.40 & 3.50 & 3.40 & 3.00 & \textbf{3.83} \\
    No inconsistencies& 3.00 & 3.17 & 2.80 & 3.00 & 3.00 & 2.60 & \textbf{3.33} \\
    Learn to understand& 3.55 & 3.50 & 3.60 & 3.50 & 3.60 & 3.20 & \textbf{3.83} \\
    Needs references& 3.73 & 3.67 & 3.80 & 3.50 & 4.00 & 3.60 & \textbf{3.83} \\
    Efficient& 3.45 & \textbf{3.67} & 3.20 & \textbf{3.67} & 3.20 & 3.40 & 3.50 \\
    \textbf{Overall score} & 0.68 & 0.68 & 0.67 & 0.68 & 0.67 & 0.64 & \textbf{0.71} \\ \hline
    \end{tabular}
    \label{tab:skincon_scs}
\end{table}

As previously discussed. We expect participants to explore a model's sensitivity to concept values. In the lay-person study accurate model we observe the clearest sign of this. In particular with concept recall where recall initially falls before increasing from 4 interventions. This shows participants appear to be exploring the concept speck before correcting concept values and making a move.

When combining all test-time intervention results, it becomes clear that \acp{CBM} are mostly not aligned to the concepts participants are adjusting. Although interventions often make concept vectors more accurate, task accuracy does not reflect this improvement. This aligns with the findings in \cite{Barker2023SelectiveCM}.

\subsection{System Causability Scale}

We used the \ac{SCS} \cite{system_causability_scale} to get a subjective rating of explanation suitability with expert study results presented in Table~\ref{tab:skincon_scs}. The overall score, computed as the average of participants' summed responses normalised by the maximum possible score. This score is between 0 and 1 where 0.68 indicates an average response \cite{system_causability_scale}. Almost all overall scores are either 0.68 or slightly below. The sub-section of participants who's overall score exceeded this are participants who selected ``strongly agree'' as their experience at classifying skin diseases in the demographic survey, with a score of 0.71. 

\begin{table}[ht!]
\centering
\caption{Lay-person study Likert scores for \acs{SCS} questions.}
\begin{tabular}{l|c|c|c|c|c|c|c|c|c|c|c}
\hline
Question & \rotatebox{90}{All Participants} & \rotatebox{90}{\acs{Acc}-\acs{NoExp}} & \rotatebox{90}{\acs{Inacc}-\acs{NoExp}} & \rotatebox{90}{\acs{Acc}-\acs{CExp}} & \rotatebox{90}{\acs{Inacc}-\acs{CExp}} & \rotatebox{90}{\acs{Acc}-\acs{CExp+Int}} & \rotatebox{90}{\acs{Inacc}-\acs{CExp+Int}} & \rotatebox{90}{\acs{Acc}-\acs{CExp+Int+SMap}} & \rotatebox{90}{\acs{Inacc}-\acs{CExp+Int+SMap}} & \rotatebox{90}{\makecell[tl]{\acs{CExp+Int}-\acs{WithInt} and \\\acs{CExp+Int+SMap}-\acs{WithInt}}} & \rotatebox{90}{\makecell[tl]{\acs{CExp+Int}-\acs{NoInt} and \\\acs{CExp+Int+SMap}-\acs{NoInt}}} \\ \hline
Factors in data & 3.39 & 3.08 & 2.85 & \textbf{3.92} & 3.69 & 3.62 & 3.50 & 3.23 & 3.25 & 2.87 & 3.25 \\
Understood & 4.18 & 4.08 & 4.15 & \textbf{4.31} & \textbf{4.31} & 4.23 & 4.17 & 4.08 & 4.08 & 4.13 & 4.08 \\
\makecell[tl]{Change detail level} & 2.91 & 2.69 & 2.31 & 2.77 & 2.38 & 3.38 & \textbf{3.50} & 3.31 & 3.00 & 3.13 & 3.00 \\
Need support & 3.77 & 3.92 & 3.92 & 3.77 & 3.62 & 3.54 & \textbf{4.17} & 3.54 & 3.75 & 3.61 & 3.75 \\
\makecell[tl]{Understanding causality} & 3.51 & 3.15 & 3.38 & 3.62 & 3.46 & 3.31 & 3.75 & 3.46 & \textbf{4.00} & 3.43 & \textbf{4.00} \\
\makecell[tl]{Use with knowledge} & 3.99 & 4.15 & 3.92 & 4.23 & 3.62 & 3.92 & 3.92 & 3.85 & \textbf{4.33} & 3.91 & \textbf{4.33} \\
\makecell[tl]{No inconsistencies} & 3.59 & 3.77 & 3.38 & \textbf{4.15} & 3.54 & 3.54 & 3.08 & 3.77 & 3.42 & 2.96 & 3.42 \\
\makecell[tl]{Learn to understand} & 4.06 & \textbf{4.31} & 4.15 & 4.00 & 4.15 & 3.92 & 4.00 & 3.92 & 4.00 & 3.74 & 4.00 \\
Needs references & 3.64 & \textbf{4.00} & 3.77 & 3.85 & 3.46 & 3.54 & 3.33 & 3.54 & 3.58 & 3.13 & 3.58 \\
Efficient & 4.24 & 4.08 & 3.85 & 4.15 & \textbf{4.62} & 4.15 & 4.42 & 4.08 & 4.58 & 4.26 & 4.58 \\
\makecell[tl]{\textbf{Overall score}} & 0.75 & 0.74 & 0.71 & \textbf{0.78} & 0.74 & 0.74 & 0.76 & 0.74 & 0.76 & 0.70 & 0.76 \\ \hline
\end{tabular}
\label{tab:blackjack_scs}
\end{table}

For individual questions, most averaged to be between high 2 and high 3 (Likart options ``disagree'' and ``neutral''). A few questions stand out. Starting with \textit{change detail level} (\textit{I could change the level of detail on demand}) was rated higher for \acs{CExp+Int} participants, \acs{WithInt} participants, and participants who self-rated their experience at skin disease identification as ``strongly agree'' (of which 57\% performed interventions). This suggests that if participants perform interventions they understand the information it provides.

\textit{Need support} (\textit{I did not need support to understand the explanations}) averaged to 4 for both \acs{CExp+Int+SMap} participants and participants who answered their skin disease identification experience as ``agree''. For \acs{CExp+Int+SMap} participants these results indicates the potential benefit saliency maps provide to help participants interpret the model's concept predictions. For skin experience agree participants, 60\% of which used interventions (with two of these participants performing almost 60 interventions), suggests their increased interaction with the model improved their understanding of the model.

Finally, \textit{efficient} (\textit{I received the explanations in a timely and efficient manner}) was also consistently rated slightly over 3. \acs{WithInt} participants answered this question with a slightly higher score than \acs{NoInt} participants.

Responses from the \ac{SCS} questions for the lay-person study are shown in Table~\ref{tab:blackjack_scs}. All overall scores are 0.70 or above. The highest score was 0.78 for \acs{Acc}-\acs{CExp} participants. Surprisingly, \acs{Acc}-\acs{NoExp} participants score was 0.74 which matches some participant groups who had access to concepts and interventions. As each participant only answered questions for one version of the explanations instead of ranking each, this may be attributed to the similar scores.

The \ac{SCS} results suggest that incorporating concepts improves participants’ understanding of causality as shown by the questions \textit{Understanding causality} (\textit{I found the explanations helped me to understand causality}) scoring higher with the inclusion of explanation techniques, although the differences between participant groups was small. Additionally, These results varied between studies where the expert study answered this question inline with the lay-person study with no model explanations. Model outputs were generally well understood as shown by the results for \textit{Understood} (\textit{I understood the explanations within the context of my work.}), and \textit{learn to understand} (\textit{I think that most people would learn to understand the explanations very quickly}).

Overall, while model outputs were generally well understood, reflected in the scores for Understood ("I understood the explanations within the context of my work") and Learn to understand ("I think that most people would learn to understand the explanations very quickly"), there are indications of a possible mismatch between human and machine decision-making. \textit{Understood}, \textit{Learn to understand}, \textit{Use with knowledge}, \textit{No inconsistencies} ("I did not find inconsistencies between explanations"), and \textit{Need references} ("I did not need more references in the explanations: e.g., medical guidelines, regulations") were all scored lower if participants made interventions. Although \textit{no inconsistencies} was also low for the inaccurate model it also shows interventions may be causing confusion over how they relate to task predictions or how they are used instead of aiding in completing the task.

\section{Discussion} \label{discussion}

Before beginning the discussion, we have detailed several limitations in our studies. The expert study had a small sample size, which may limit the generalisability of our findings. To address this, we conducted a larger lay-person study and drew parallels between the two to provide additional context. Additionally, participants in the expert study lacked access to patient history, high-quality diagnostic images, and multiple images of each sample, which may be expected in clinical settings. To mitigate this, we simplified the task to distinguish between ``malignant melanoma'' and ``seborrhoeic keratosis'' which have clear visual differences. Finally, in the lay-person study, participants played Blackjack, meaning game success depended partly on luck, though our evaluation focused on optimal moves rather than overall game outcomes.

From our human studies evaluating \acp{CBM}, we observed mixed results regarding interpretability and task performance. While our findings reinforce \acp{CBM} interpretability with participants who utilised concepts and interventions to explore the concept space and inspect task predictions, task accuracy improvements were inconsistent. Notably, while concept accuracy increased, task accuracy mostly decreased.

\subsection{Do Test-time Interventions Improve Human-machine Task and Concept Accuracy?}

Test-time interventions found mixed results across models. In most cases, task accuracy with interventions matched or underperformed the model's accuracy with no interventions, with further declines in task accuracy as the number of interventions increased. The only notable increases in model task accuracy was seen in the lay-person study with the inaccurate model. Following our test-time intervention results, it suggests interventions have the risk of leading to decreased task accuracy if humans follow model task predictions after interventions are performed.

For concept accuracy, interventions increased accuracy. Despite this not being consistent across all models, decreases in some situations (e.g. accurate blackjack model) were expected to account for participants learning the model's sensitivity to concepts.

Similar to \cite{Barker2023SelectiveCM}, our findings suggest that \acp{CBM} task predictions may use different concepts than humans use. Future research should explore methods to align \ac{CBM} decision-making with human decision-making.

\subsection{Do Interventions Increase the Interpretability of Concept Bottleneck Models?}

In the expert study, interventions were almost evenly split between error correction and feature adjustments. Further, intervention decreased over time, suggesting that participants relied on interventions less as they developed a mental model of the model's behaviour. This aligns with the idea that \acp{CBM} improve interpretability.

Saliency maps decreased the number of interventions performed suggesting they provide sufficient insight into the model’s behaviour. However, participants also reported they placed little weight on the model's predictions, implying that participants preferred their own intuition than relying on the model.

Regarding the lay-person study, participants using the accurate models primarily performed feature adjustments, with nearly 75\% of interventions involving changes to concept presence. Combined with the eventual increase in concept recall, this indicates participants used the interpretability of concepts to improve their understanding of the model. More interventions, including reversals, were performed with the inaccurate model.

While we observe a decline in interventions over time in both studies, part of this decline may be attributed to the novelty of interventions, with engagement naturally decreasing as participants became more familiar with the task. Although we cannot entirely rule out this effect, the fact that the decline is not uniform, particularly in the lay-person study, where interventions remained higher when concepts were incorrectly predicted suggests that participants were not merely losing interest but actively leveraging interventions to improve their understanding of the model.

Our findings support the claim that \acp{CBM} improve interpretability by allowing users to interactively query and adjust concept predictions. However, we have identified this process is limiting as humans are required to seek explanations and iteratively probe the model’s concept sensitivity, which may not be practical or obvious for all users or applications. Further, our study does not look at the role of the interface in engaging participants to interact with the model. We suggest future research should look at the delivery of concept explanations to ensure they are efficiently delivered.

\subsection{Are Concept Bottleneck Models Trusted?}

We chose to use alignment as a proxy for trust \cite{10316181}. In the expert study, \acs{NoInt} participants aligned to the model's predictions 81\% of the time, which is 11\% higher than the model's accuracy on the samples in the study suggesting over-trust. In contrast, \acs{WithInt} participants were aligned to the model's initial task prediction 66\% of the time, 4\% lower than the model's accuracy. Alignment then increased by almost 13\% after interventions. In addition, accuracy was higher for \acs{WithInt} participants compared to \acs{NoInt} participants. This shows interventions increased trust, which itself was better justified compared to participants who did not use interventions.

For the lay-person study, alignment was lower than the model’s accuracy before interventions. When participants used interventions, alignment increased significantly across all participant groups. However, this increase in alignment only increased task accuracy for participants using the inaccurate model. This shows the potential for interventions to lead to over-trust. In addition, providing just concept explanations lead to an increase in alignment while also increasing task accuracy.

The trends of alignment and joint task accuracy are conflicting between the studies. We hypothesis this is because in the expert study the model outputs were used purely as a second opinion as the participants would have sufficient expertise in the task domain. As this is not guaranteed in the lay-person study we believe participants may follow the model if they are unsure themselves. This is a concerning point if these models are deployed in situations where humans are not domain experts.

\section{Conclusion}

In this paper, we ran the first studies to evaluate how humans use \acp{CBM} in a collaborative setting. We focused on how concepts are interacted with and the interpretability of these models. In particular, we evaluate (1) if concept interventions increase the model's task accuracy, (2) do concept interventions increase model interpretability, and (3) are \acp{CBM} are trusted. We find \acp{CBM} do not translate to increased model task accuracy in a human-machine setting, but this model architecture and other \acp{CM} are shown to increase both the interpretability and trust with the model's task label predictions.

We drew three main conclusions from our studies:

Firstly, interventions significantly improved concept accuracy but had limited impact on task accuracy. This suggests a misalignment between the concepts humans use and the concepts the models use to label samples. Addressing this misalignment is critical to improving the effectiveness of \acp{CM} human-machine teams.

Next, we show the initial promise of interpretability from high-level concepts and interpretability is upheld with \acp{CM}. However, as this required participants to engage in interventions, we highlight a need for \acp{CBM} to present their decision-making process proactively, reducing the cognitive effort required from humans. In addition, much of the interpretability can be provided by just providing concept predictions.

Finally, using alignment as a proxy for trust, we found that interventions led to higher trust. This did not always lead to increased task accuracy and, as shown in the lay-person study, can result in overtrust. This highlights the importance of interpretable models that are evaluated with human participants to enable the creation of trust that is suitably applied to a model.

\begin{credits}
\subsubsection{\ackname} This research was funded by the UK Engineering and Physical Sciences Research Council (EPSRC) and IBM UK via an Industrial CASE (ICASE) award.

\subsubsection{\discintname}
The authors have no competing interests to declare that are relevant to the content of this article.
\end{credits}

\bibliographystyle{splncs04}
\bibliography{bibliography}

\end{document}